\documentstyle{l-aa}
\input epsf

\begin{document}
\thesaurus{08               %Diffuse matter...
                 (02.08.1;  %Hydrodynamics
                  03.13.1;  %Method:analytical
                  09.19.2;  %ISM:supernova remnants
                  13.25.4   %X-rays:interstellar
                  )
}
\title{Evolution of supernova remnants in the interstellar medium
	with a large-scale density gradient}
\subtitle{I. General properties of the morphological evolution 
	and X-ray emission}
\author{B.~Hnatyk
\and O.~Petruk}
\institute{Institute for Applied Problems in Mechanics and Mathematics 
           NAS of Ukraine, 3-b Naukova St., Lviv 290601, Ukraine\\
           }
\offprints{B. Hnatyk (Hnatyk@lms.lviv.ua)}
\date{Received ...; accepted ...}
\maketitle
\markboth{B.~Hnatyk and O.~Petruk: Evolution of SNRs in the ISM with 
	a large-scale density gradient}{}
%%%%%%%%%%%%%%%%%%%%%%%%%%%%%%%%%%%%%%%%%%%%%%%%%%%%%%%%%%%%%%%%%%%%%%%%%
%%%%=====================Abstract====================================%%%%
%%%%%%%%%%%%%%%%%%%%%%%%%%%%%%%%%%%%%%%%%%%%%%%%%%%%%%%%%%%%%%%%%%%%%%%%%

\begin{abstract}

The large-scale gradient of the interstellar medium (ISM) density 
distribution essentially affects the evolution of Supernova remnants (SNRs).  
In a non-uniform 
ISM, the shape of SNR becomes essentially non-spherical, and distributions of 
gas parameters inside the remnant become strongly anisotropic. The well-known 
self-similar Sedov solutions may not be applied to modelling such 
non-spherical objects.  Therefore we propose a new approximate analytical 
method for full hydrodynamical description of 3D point-like explosions in 
non-uniform media with arbitrary density distribution.  \par

On the basis of this method, we investigate the general properties of 
evolution of 2D non-spherical adiabatic SNRs in ISM with large-scale density 
gradient.  \par

    It is shown that the real shape of adiabatic SNR becomes more non-spherical 
with age, but the visible shape remains close to spherical even for 
strong real asymmetry (ratio of maximal to minimal shock radii) and 
surface brightness contrast.  It is shown also that the values of 
the parameters of X-ray radiation from the entire SNR (luminosity, 
spectral index) are close to those in the Sedov case with the same 
initial data. However, the surface distribution of the X-ray emission 
parameters is very sensitive to the initial density distribution around 
the SN progenitor. Therefore, 
the X-ray maps give important information about physical 
conditions inside and outside of the non-spherical SNR. \par 

\end{abstract}

%%%%%%%%%%%%%%%%%%%%%%%%%%%%%%%%%%%%%%%%%%%%%%%%%%%%%%%%%%%%%%%%%%%%%%%%%
%%%%========================Text=====================================%%%%
%%%%%%%%%%%%%%%%%%%%%%%%%%%%%%%%%%%%%%%%%%%%%%%%%%%%%%%%%%%%%%%%%%%%%%%%%

\medskip

%%%%%%%%%%%%%%%%%%%%%%%%%%%%%%%%%%%%%%%%%%%%%%%%%%%%%%%%%%%%%%%%%%%%%%%%%
%%%%======================Section I==================================%%%%
%%%%%%%%%%%%%%%%%%%%%%%%%%%%%%%%%%%%%%%%%%%%%%%%%%%%%%%%%%%%%%%%%%%%%%%%%

\section{Introduction}

   The investigation of Supernova remnants (SNRs) gives important
information on the physics of Supernova (SN) explosions, on the properties 
of the surrounding
interstellar medium (ISM) and on shock wave physics.  The majority of
galactic SNRs are in the adiabatic stage of evolution 
(Lozinskaya \cite{Lozins}). 
If the density of the ISM is uniform, their hydrodynamics are
well described by the self-similar Sedov solution (Sedov \cite{Sedov}, 
Shklovskiy \cite{Shklovskyj62}). The typical values of plasma 
temperatures in SNRs are
$T\simeq 10^6$ to $10^8\ {\rm K}$ and, therefore, SNRs radiate mainly 
in the X-rays. The spectral characteristics of the equilibrium X-ray 
emission for a plasma
typical SNR abundances was calculated by Shapiro \& Moore
(\cite{Shapiro-Moore}), Raymond \& Smith (\cite{Raym-Smith77}), Shull
(\cite{Shull81}) and Gaets \& Salpeter (\cite{Gaets}). \par

   But the real situation is more complicated. In many SNRs, the plasma 
is in nonequilibrium ionization (NEI). Often there is no thermal equilibrium 
between the electrons and the ions. The physical conditions in the inner 
parts of SNR may be modified by electron thermal conductivity (Itoh
\cite{Itoh77}, Cox \& Anderson \cite{Cox-And}, Hamilton et al.
\cite{Hamilton}, Jerius \& Teske \cite{Jer-Teske}, 
Borkowski et al.  \cite{Bork-Sar-Blond94}, Bocchino et al.
\cite{Bocc-Magg-Scio97}).  There are also other effects which affect 
the plasma emission in SNR, but their influence is small: 
resonant scattering, diffusion etc. 
(Raymond \& Brickhouse \cite{Raym-Brickh} and references there). 
The Sedov solution may also be modified by the presence
of small-scale cloudlets in the ISM (Bychkov \& Pikelner \cite{Bychk-Pic75}, 
McKee \& Cowie \cite{McKee-Cow75}, Sgro \cite{Sgro}, 
White \& Long \cite{White-Long91}).  \par

\begin{figure}
%\picplace{7.5 cm}
\epsfysize=7truecm
\epsfbox{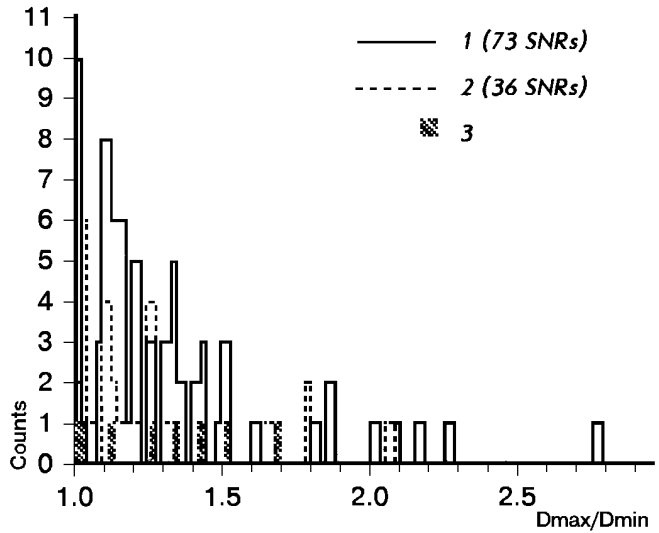}
\caption[]{
Distribution of ratios of maximal to minimal visual diameters of
SNRs from 1 - radio observations (Whiteoak \& Green
\cite{MOST-catalog}) and from 2 - an X-ray catalogue (Seward
\cite{Seward-catalog}). The young SNRs, which are
expecting to be in the free expansion stage, are 
separated - 3 (Lozinskaya \cite{Lozins}, Seward
\cite{Seward-catalog}, Greiner et al.  \cite{Gr-E-A}, Aslanjan
\cite{Aslanjan96}, Stankevich \cite{Stankevich96}).  
           } 
\label{gistogr}
\end{figure}

  Practically the all above-mentioned investigations 
assumed spherical symmetry of SNRs, as the result of their evolution in {\it
uniform} ISM. However, the observations show predominantly nonspherical
shapes (Seward \cite{Seward-catalog}, Whiteoak \& Green
\cite{MOST-catalog}). For example Kesteven \& Caswell (\cite{Kesteven}) 
and Bisnovatyi-Kogan et al.  (\cite{BK-L-S}) define a separate class of
barrel-like SNRs. We have carried out the analysis of visual anisotropy
for SNRs from the catalogues of Seward (\cite{Seward-catalog}) 
and Whiteoak \& Green (\cite{MOST-catalog}) (Fig.~\ref{gistogr}). 
This figure shows that typical values
of the visual anisotropy is usually $d_{\rm max}/d_{\rm min}=1$ to $2$. \par

   Many SNRs with nearly spherical visual shapes have an 
anisotropic distribution of surface brightness (e.g.  Kepler
SNR, Cygnus Loop, RCW86 etc.).\par

Therefore, truly spherical SNRs are considerably rarer than believed. \par

Non-spherical SNRs (NSNRs) may be created by an anisotropic SN explosion 
(e.g.  Bisnovatyi-Kogan \cite{BK}). Non-spherical shapes of SNRs in the free 
expansion stage (Fig.~\ref{gistogr}) should be mainly produced in this way. 
Another important reason for the non-sphericity of 
adiabatic SNRs can be a non-uniform 
density distribution of the ISM or the large-scale magnetic field. 
In such cases, self-similar solutions cannot be used, while direct
numerical calculations of the problem are difficult because of the
complications of 3D hydrodynamical modelling SNR
evolution multiplies by the complications of nonequilibrium
describing of gas element evolution.  Therefore, at  present only a few
simplified models have been built (Tenorio-Tagle et al.
\cite{TT-R-Yo}, Bodenheimer et al.  \cite{B-Yo-TT}, Claas \& Smith
\cite{Claas-Smith89}, Bocchino et al.  \cite{Bocc-Magg-Scio97}).\par

   Real possibility to perform an investigation of the evolution of
non-spherical SNRs bases on approximate methods for 
hydrodynamics. The thin-layer (Kompaneets
\cite{Komp}) approximation for the calculation of SNR shape is widely
used (Lozinskaja \cite{Lozins}, Bisnovatyi-Kogan \&
Silich \cite{BK-Syl} and references there).  But this 
approximation has low accuracy for the adiabatic
stage in a nonuniform medium and does not allow to calculate
the behaviour of the gas inside the SNR (Hnatyk \cite{Hn87}, 
Hnatyk \& Petruk \cite{Hn-Pet-96}).  Hnatyk (\cite {Hn87}) has 
shown that for our 
probleme it is more promising to develop approximate
methods under a sector approximation (Laumbach \& Probstein
\cite{L-P}), which allows to calculate both the SNR shape and the gas
characteristics. In the work of Hnatyk \& Petruk (\cite{Hn-Pet-96})
a new approximate analytical method for the complete
hydrodynamical description of a point explosion in a medium with
an arbitrary regular but smooth density distribution was proposed . 
It combines the 
advantages of the two above-mentioned methods and, therefore allows to
calculate the hydrodynamical aspects of non-sherical SNR evolution 
with high enough accuracy in a short computing time.\par

%%%%=== Fig ===%%%%%%%%%%%%%%%%%%%%%%%%%%%%%%%%%%%%%%%%%%%%%%%%%%%%%%%%%%%%%%%
%%%%%%%%%%%%%%%%%%%%%%%%%%%%%%%%%%%%%%%%%%%%%%%%%%%%%%%%%%%%%%%%%%%%%%%%%%%%%%

  We use this method here, but we limit ourselves to the case of 
equilibrium emissivity. Non-equilibrium effects will be considered 
in a further paper. \par

%%%%%%%%%%%%%%%%%%%%%%%%%%%%%%%%%%%%%%%%%%%%%%%%%%%%%%%%%%%%%%%%%%%%%%%%%
%%%%======================Section II=================================%%%%
%%%%%%%%%%%%%%%%%%%%%%%%%%%%%%%%%%%%%%%%%%%%%%%%%%%%%%%%%%%%%%%%%%%%%%%%%
\section {Hydrodynamical modelling}

  As proposed by Hnatyk \& Petruk (\cite{Hn-Pet-96}) the hydrodynamical 
description includes two steps: the calculation of shock front dynamics 
(shape of SNR) and the calculation of the state of the plasma inside the 
SNR.

\subsection{Calculation of shock front motion}

   Klimishin \& Hnatyk (\cite{Klym-Hn81}), Hnatyk (\cite{Hn87}) have shown
that the motion of a strong one-dimensional adiabatic shock wave in a 
medium with an arbitrary distribution of density is described with high 
accuracy by the approximate formula:
\begin{equation} 
\label{apr-form} 
{dR \over dt} = D(R) = {\rm const} \cdot
(\rho^{\rm o}(R)\cdot R^{N+1})^{-k}
\end{equation}
where
\begin{equation}
\label{koef-apr-form}
k=\left\{\matrix {1/2, & m(R)\le N+1 &\cr
                  1/5, & m(R) > N+1  &\cr}\right.\ ,
\end{equation}
$R$ is the distance from the explosion, $\rho^{o}(R)$ is the 
initial density distribution of the surrounding medium; $m(R)=-d\ln
(\rho^{o}(R)) / d\ln R$; $N=0, 1, 2$ for a plane, cylindrical and
spherical shock, respectively. \par

Formulae (\ref{apr-form})-(\ref{koef-apr-form}) generalize two 
basic features of shock motion in non-uniform media:\par

   - {\em a deceleration} in a medium with
increasing $(m(R)<0)$, constant $(m(R)=0)$ or slowly decreasing
density $(0<m(R)\le N+1)$; then the parameter $k$ is close to $k=1/2$;
\par

   - {\em an acceleration} if the density is decreasing 
fastly enough as $m(R)>N+1$; then the parameter $k$ is close to $k=1/5$.\par

   Since in all realistic cases $m(0)=0$, i.e. the initial stage of the 
motion of the shock from a point explosion is always described by the 
self-similar Sedov solution for a {\em uniform} medium, for the deceleration 
stage of shock motion from eq. (\ref{apr-form})-(\ref{koef-apr-form}) 
we have:
\begin{equation}
\label{D-af-ietap}
\begin{array}{l}
D(R)\approx D_{\rm D}(R) = \\
\qquad\quad ={\displaystyle {2\over 3+N}\cdot \biggl({E_{o} \over \alpha_{\rm A}
(N,\gamma)\cdot \rho^{\rm o}(R)}\biggr)^{1/2}\cdot R^{-(N+1)/2}},
\end{array}
\end{equation}
where $\alpha_{\rm A} (N,\gamma)$ is the self-similar constant for a uniform 
medium ($\alpha_{\rm A}(0,\ 5/3)=0.6029$, $\alpha_{\rm A} (2,\ 5/3)=0.4936$), 
$E_{o}$ is the energy of explosion. \par

      If the density distribution is such that a region of acceleration with 
$m(R)>N+1$ exists and begins at some distance $R_1$ where $m(R_1)=N+1$,
then an approximate formula for shock velocity is: 
\begin{equation} 
\label{D-af-iietap} 
D_{\rm A}(R,R_1)\approx D_{\rm D}(R_1)
\cdot \biggl( {\rho^{\rm o}(R_1)\cdot R_1^{N+1}\over\rho^{\rm o}(R)\cdot R^{N+1}}
\biggr)^{1/5}
\end{equation}
Here, $D_{\rm D}(R_1)$ is determined from (\ref{D-af-ietap}).\par

   If the density distribution is such that there is a transition from a 
region
of acceleration with $m(R)>N+1$ into a region of deceleration where $m(R)\le
N+1$, then another feature appears.  At the beginning, 
the shock deceleration will be analogous to the motion of an external
(forward) shock as described by Nadyezhyn (\cite{Nadyezh85}) and Chevalier
(\cite{Schevalier}).  The external shock decelerates with 
$k\approx 1/5$ and, when its velocity equals velocity $D_{\rm D}(R)$ from
(\ref{D-af-ietap}) the further deceleration is described by $k\approx
1/2$.\par

Therefore, a general formula for the shock velocity which takes into 
account the three cases is (Hnatyk \cite{Hn87}):
\begin{equation}
\label{D-af-total}
\begin{array}{l}
D(R) = \\
\\
\left\{\matrix {
D_{\rm D}(R),&0 \le R \le R_1&\cr
min[D_{\rm A}(R;R_{2n-1}); D_{\rm D}(R )],&R_{2n-1} < R < R_{2n+1}&\cr}\right.
\!\!\!\!\!\!\!\!\!\!\!\!
\end{array}
\end{equation}
where $n=1,2,3...$; $R_n$ are zeros of the function
$m(R)-(N+1)$ (the points of changing of regimes of shock motion) 
in increasing order of $R$ in the interval $(0, \infty)$. \par

%%%%=== Fig ===%%%%%%%%%%%%%%%%%%%%%%%%%%%%%%%%%%%%%%%%%%%%%%%%%%%%%%%%%%%%%%%
\begin{figure}[t]
%\picplace{12 cm}
\epsfysize=12truecm
\epsfbox{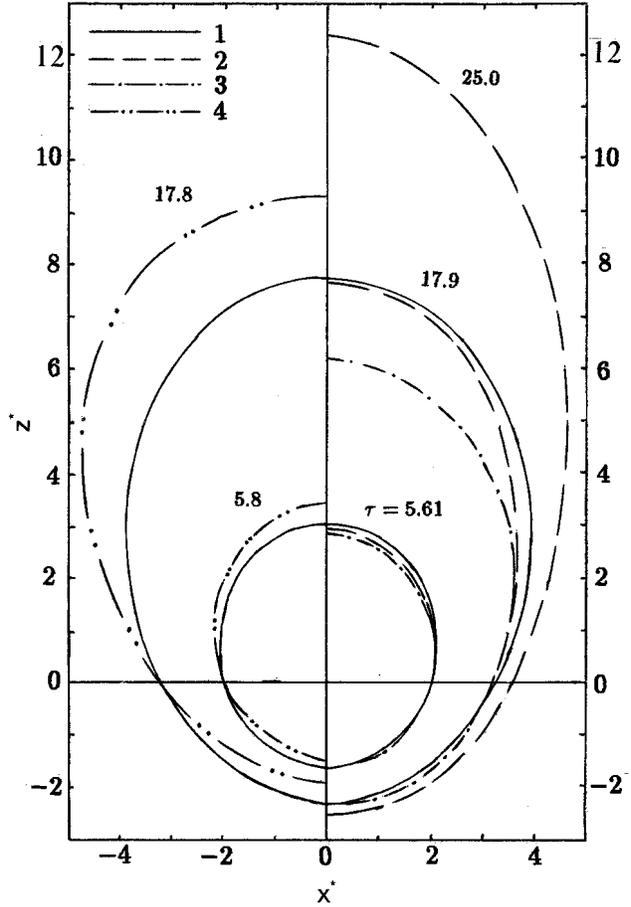}
\caption[]{
Profiles of a shock front in a 
non-uniform exponential medium Eq.~(\ref{exp-density}) as a function of 
dimentionless time $\tau$ (see text). The coordinates are non-dimensional 
as defined in the text. Line 1 is the result of the numerical 2D 
calculation; line 2 is the approximate law (\ref{D-af-total}); 
line 3 is the result of numerical calculation in the sector approximation; 
line 4 is the result of the thin layer approximation.
           }
\label{test-shape-exp}
\end{figure}
%%%%%%%%%%%%%%%%%%%%%%%%%%%%%%%%%%%%%%%%%%%%%%%%%%%%%%%%%%%%%%%%%%%%%%%%%%%%%%

%%%%=== Fig ===%%%%%%%%%%%%%%%%%%%%%%%%%%%%%%%%%%%%%%%%%%%%%%%%%%%%%%%%%%%%%%%
\begin{figure*}[t]
%\picplace{9 cm}
\epsfysize=7.8truecm
\epsfbox{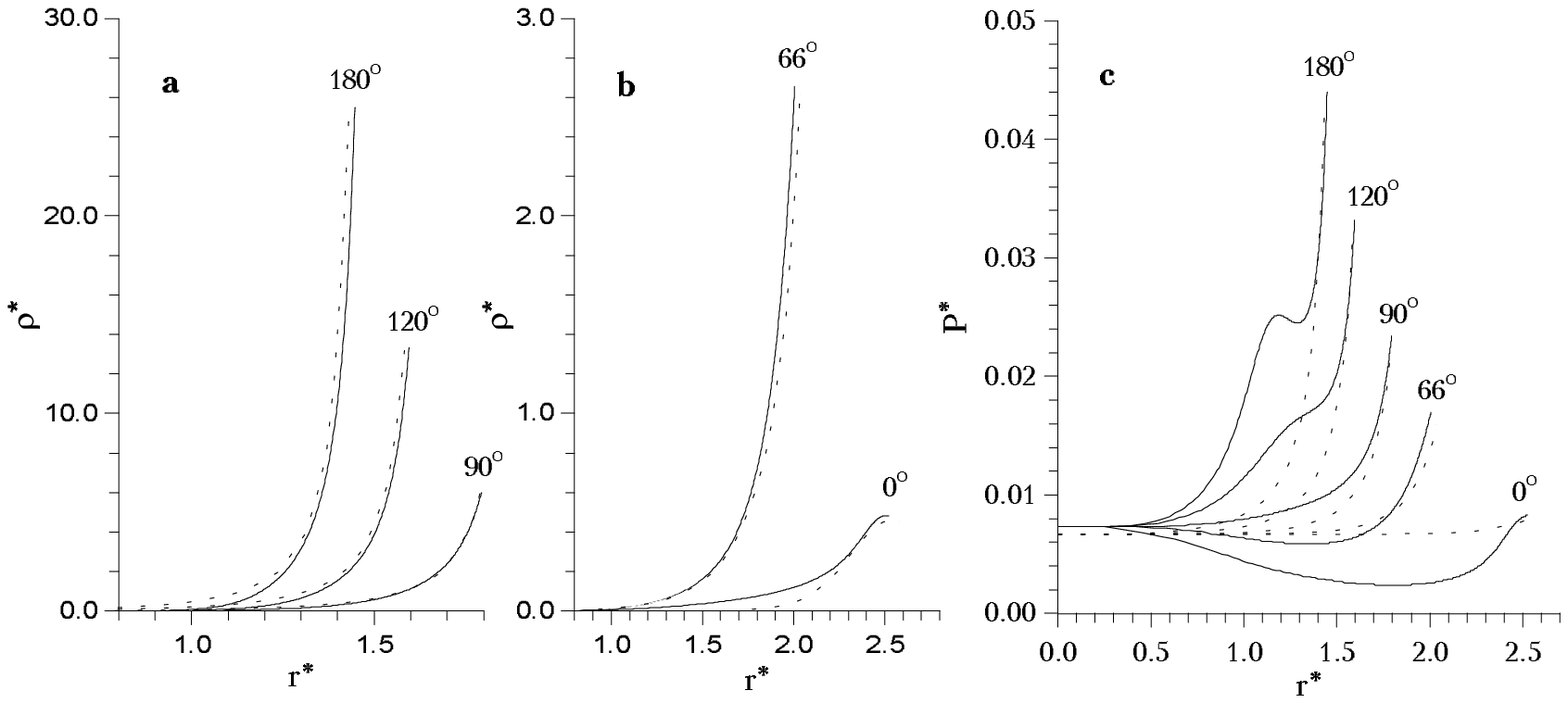}
\caption[]{
Distributions of density ({\bf a, b}) and pressure ({\bf c}) inside an SNR 
expanding in a flat exponential medium Eq.~(\ref{exp-density}) for different 
directions from the place of explosion at the dimensionless time 
$\tau=4.268.$  The solid lines correspond to the present 
method, the dashed lines are the result of the direct 2D numerical 
calculations of Kestenboim et al. (\cite{Kestenb}). Here $\gamma=1.4.$ 
           }
\label{test-profinside-exp}
\end{figure*}
%%%%%%%%%%%%%%%%%%%%%%%%%%%%%%%%%%%%%%%%%%%%%%%%%%%%%%%%%%%%%%%%%%%%%%%%%%%%%%

   It must be emphasized that the majority of models interesting for
astrophysics are described more simply by formulae 
(\ref{D-af-ietap})-(\ref{D-af-iietap}).  \par

The equation of the trajectory of the shock motion is
\begin{equation}
\label{traectorUCh}
t = \int\limits_0^R {dR \over D(R)}.
\end{equation}

   When the density distribution departs from spherical symmetry, the
calculation should be carried out by sectors. The 
3D region is divided into the necessary quantity of sectors and
Eq.~(\ref{traectorUCh}) is integrated for each of them.
Such an approach allows also to consider an anisotropic 
energy release $E(\Omega)$ replacing $E_{\rm o}$ by 
$2\pi N E(\Omega)$ for $N=1,2$. \par

%%%%=== Fig ===%%%%%%%%%%%%%%%%%%%%%%%%%%%%%%%%%%%%%%%%%%%%%%%%%%%%%%%%%%%%%%%
\begin{figure*}[t]
%\picplace{9 cm}
\epsfysize=8.7truecm
\epsfbox{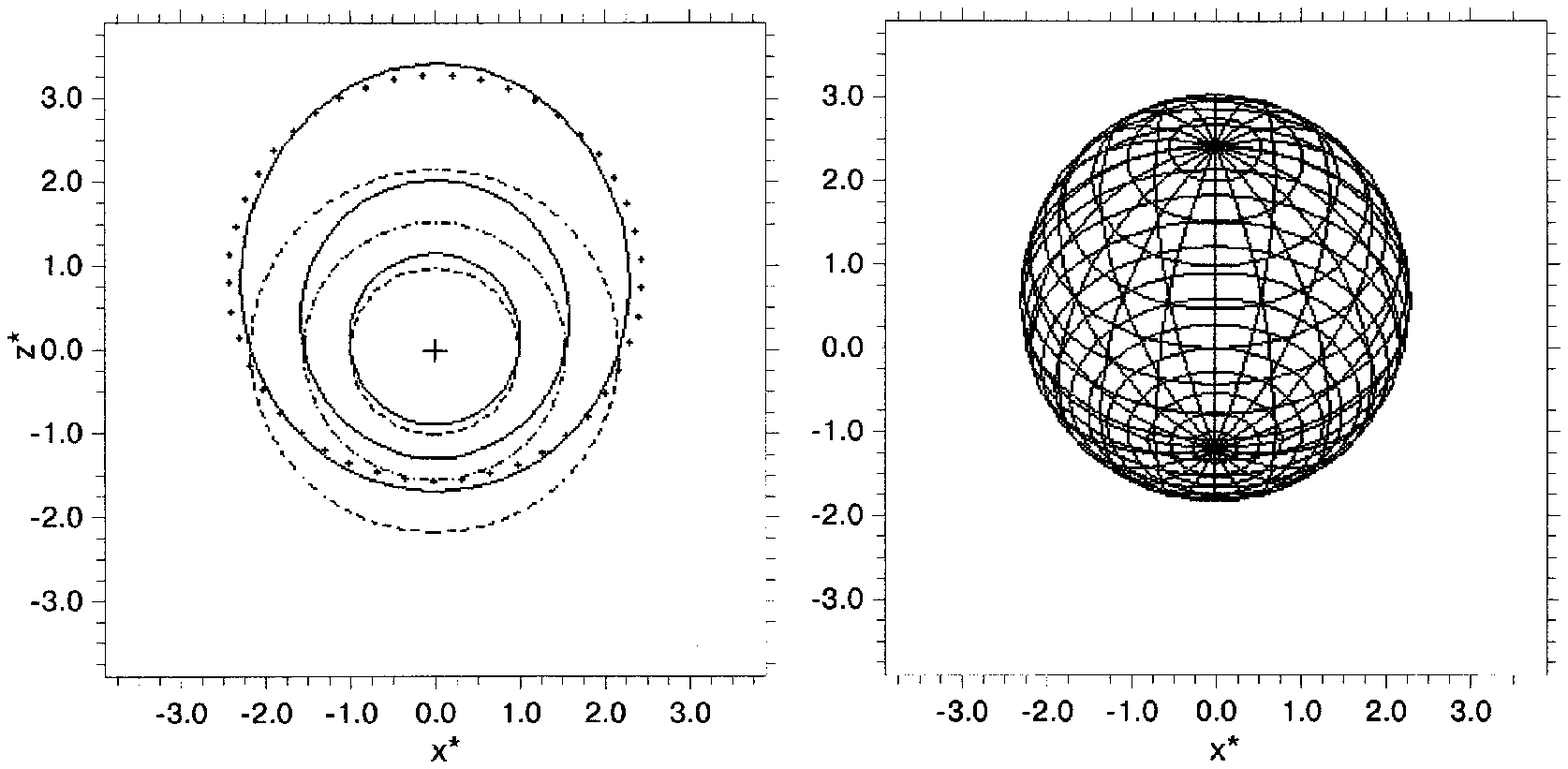}
\caption[]{
{\bf a)}~Profiles of shock front expanding in a 
exponential medium Eq.~(\ref{exp-density}) (solid lines) and in a 
uniform medium (dashed lines) for dimensionless times 
$\tau=1,\ 3,\ 7$. The dotted line is the profile fitting by a sphere
for $\tau=7$. 
{\bf b)}~The 3D~shape of the same SNR for $\tau=7$. The angle between 
the symmetry axis of the SNR and the plane of sky is $45^{\rm o}$.
           }
\label{shape-exp}
\end{figure*}
%%%%%%%%%%%%%%%%%%%%%%%%%%%%%%%%%%%%%%%%%%%%%%%%%%%%%%%%%%%%%%%%%%%%%%%%%%%%%%

\subsection{Calculation of the plasma characteristics inside SNR}

   The second step concerns the
determination of gas parameters inside the SNR. It uses on the shock motion
law by sector discribed in the previous step.  We will use the
Lagrangian coordinates $(a,t)$ ($a$ is the initial coordinate of the gas
element at $t=0$) and take as the unknown function the position
(Eulerian coordinates) of the gas element $r(a,t)$ ($0\leq r(a,t)\leq R$) at
time $t$.  The function $r(a,t)$ is expanded into series about the 
shock front and about the center of explosion. Then these two
decompositions are combined.  The coefficients of decomposition near the
shock front is determined by the motion law in each sector and
near the center of explosion assuming a zero pressure
gradient (see Appendix~\ref{App-r_o}). It is important that the value of pressure
in the central region be taken equal in all sectors contrary to
previous propositions (Laumbach \& Probstein \cite{L-P}, Hnatyk
\cite{Hn87}, Hnatyk \cite{Hn88}).  The value of pressure around
the point of explosion is determined from the condition that the ratio
$\eta$ of pressure in the center of explosion $P(0,t)$ to the average 
value of energy density inside SNR $E_{\rm o}/V_{\rm tot}(t)$ does 
not change with time and is equal to the self-similar value 
for a uniform medium: 
\begin{equation} 
\label{umova_C} 
\eta \equiv P(0,t)V_{\rm tot}(t)/E_{\rm o}~=~\eta_{\rm A}.
\end{equation}
In this way, we approximatively take into account
a redistribution of energy between the sectors in the cases of an anisotropic 
explosion and/or a non-uniform medium.  \par

The calculation of $r(a,t)$ is presented in 
Appendixes~\ref{App-eqs}-\ref{App-r_o}. 

 The other parameters (pressure $P(a,t)$, density $\rho(a,t)$, velocity
$u(a,t)$) are exactly derived from $r(a,t)$ in case of adiabatic motion of
gas behind the shock front (Hnatyk \cite{Hn87}):\newline
-- the density is obtained from the continuity equation
\begin{equation}
\label{ro(at)} 
\rho(a,t)= \rho^{\rm o}(a)\Bigl({a\over r(a,t)}\Bigr)^N
	\left({\partial r(a,t)\over\partial a}\right)^{-1};
\end{equation}
  -- the pressure is derived from the equation of adiabaticity 
$P=K\rho^{\gamma}$ or 
\begin{equation} 
\label{P(at)} 
{P(a,t)\over P(R,t)}=\biggl({\rho^{\rm o}(a)\over
 \rho^{\rm o}(R)}\biggr)^{1-\gamma} \biggl({D(a)\over
   D(R)}\biggr)^2\biggl({\rho(a,t)\over \rho(R,t)} \biggr)^{\gamma};
\end{equation}
   -- velocity directly from approximation $r=r(a,t)$
\begin{equation}
\label{u(at)}
u(a,t)={\partial r(a,t)\over \partial t}.
\end{equation}

\subsection{Testing the method}

The proposed method exactly describes the shock trajectory in self-similar 
(Sedov) cases with $m(R)={\rm const}\le N+1,$ including in particular the 
uniform ($m=0$) medium. The space distributions of the gas parameters: pressure 
$P(r,t),$ density $\rho(r,t),$ temperature $T(r,t)$ and velocity 
$u(r,t)$ inside the SNR are accurate within 3\% in the $m=0$ case 
(Hnatyk \& Petruk \cite{Hn-Pet-96}). \par

The result of the calculation for a point explosion in medium with 
an exponential density distribution 
\begin{equation} 
\label{exp-density}
\rho^{\rm o}(z)=\rho^{\rm o}(0)\cdot \exp \Bigl(-{z\over H}\Bigr),
\quad z=r \cos \theta, 
\end{equation}
where $r$ is the distance from the center of explosion, $\theta$ is the angle
between the considered direction and the direction opposite to the density
gradient, and $H$ is the scale height, are presented in 
fig.~\ref{test-shape-exp}-\ref{test-profinside-exp}.  
The simulations are carried out for
dimensionless parameters $\tau=t/t_{\rm m},$ $r^*=r/R_{\rm m},$ 
$\rho^*=\rho/\rho_{\rm m},$ $P^*=P/P_{\rm m},$  
with
$t_{\rm m}=\alpha_{\rm A}(2,\gamma)^{1/2}E_{\rm o}^{-1/2}\rho_{\rm o}^{1/2}(0)R_{\rm m}^{5/2},$ 
$R_{\rm m}=H,$ 
$\rho_{\rm m}=\rho^{\rm o}(0),$ 
and $P_{\rm m}=\alpha_{\rm A}(2,\gamma)^{-1}E_{\rm o}R_{\rm m}^{-3}.$
They allow to obtain the solution for any value 
of the initial parameters $E_{\rm o},$ $\rho^{\rm o}(0),$ 
and $H.$ \par

Fig.~\ref{test-shape-exp} presents the results of a direct 2D numerical 
calculation together with results of different approximate approaches 
(Kestenboim et al. \cite{Kestenb}). As we see, for the largest time 
$\tau=17.9,$ the proposed method is accurate within 7\%, i.e., of the 
same order of accuracy as the numerical method. Even a numerical calculation 
by sector (1D numerical calculation in each sector) yields a considerably 
lower accuracy, only about 20\%. \par

Another important quantity useful to check the accuracy the shock 
acceleration in exponential atmosphere is the breakout time $\tau_{\rm br},$ 
when shock velocity $D(\theta\!=\!0^{\rm o})\to \infty.$
The combination of the 2D numerical solutions for $\tau\le 17.9$ 
with the self-similar 
solution for $\tau>17.9$ for $\gamma=1.4$ yields $\tau_{\rm br}=29.6$ 
(Kestenboim et al. \cite{Kestenb}) while approximation (\ref{D-af-total}) 
gives $\tau_{\rm br}=30.8.$ \par

The accuracy of the density and pressure calculations in the 
proposed method is illustrated by Fig.~\ref{test-profinside-exp}.\par

Additional results of testing of the method for more complex density 
distributions, such as a density discontinuity are presented by 
Hnatyk (\cite{Hn87}). All these results reveal that the proposed method 
has high enough accuracy in all cases in the sector aproximation. 
It fails however in the case of shock interaction with a small dense 
cloud. \par

\section{SNR shapes in non-uniform media}

We now consider the role of the surrounding ISM on the evolution
and the X-ray emission of adiabatic SNR on the 
examples of media with flat exponential and spherically-symmetrical 
power-law density distributions.  \par

%%%%=== Fig ===%%%%%%%%%%%%%%%%%%%%%%%%%%%%%%%%%%%%%%%%%%%%%%%%%%%%%%%%%%%%%%%
\begin{figure*}
%\picplace{8 cm}
\epsfysize=8truecm
\epsfbox{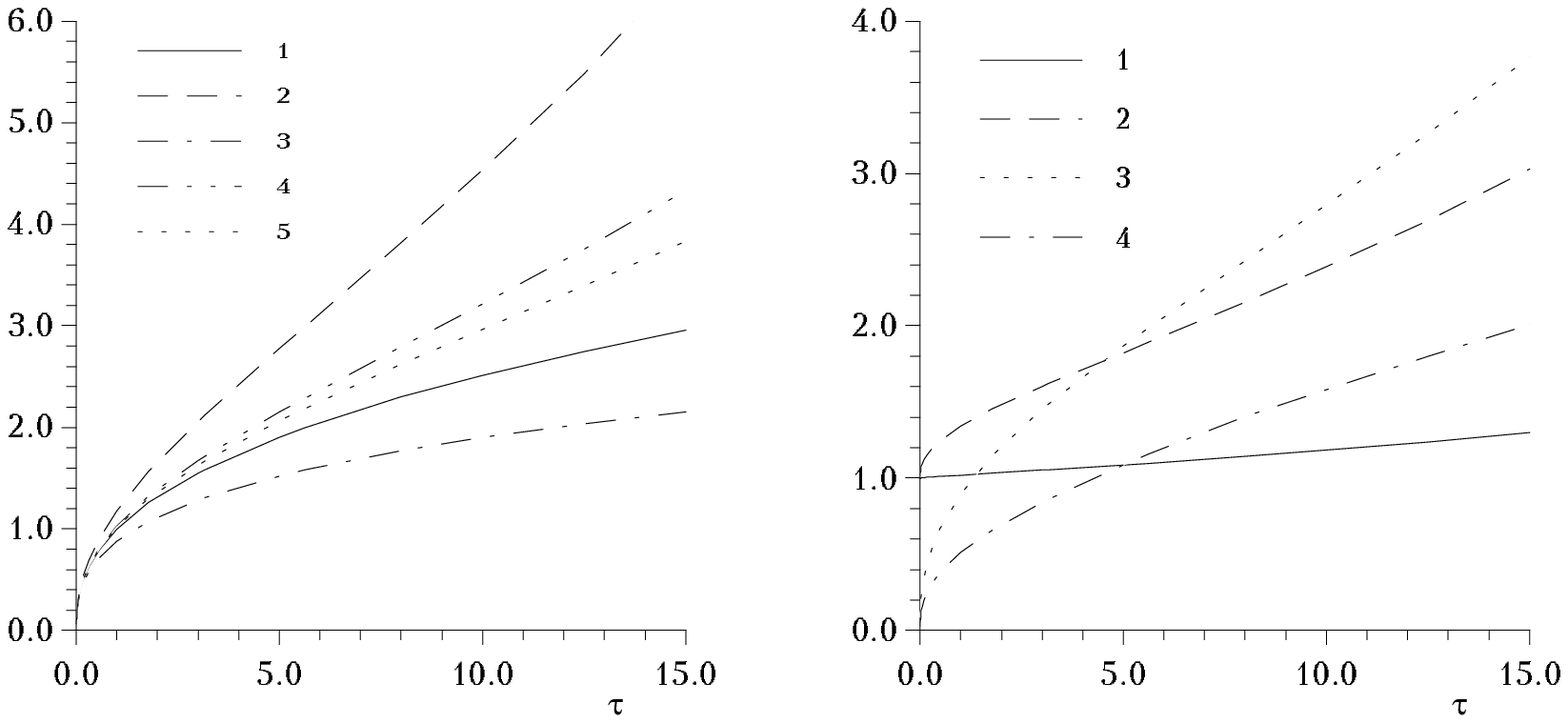}
\caption[]{
Evolution of the shape and characteristics of a shock wave for 
a point explosion in a flat exponential medium Eq.~(\ref{exp-density}).
{\bf a}:~1~--~radius of shock front $R^*_{\pi/2}$ in direction 
$\theta=90^{\rm o}$ (this equals $R_{\rm s}$ for the case a uniform medium);  
2 -- $R^*_{0}$; 
3 -- $R^*_{\pi}$; 4 -- $(R^*_{0}+R^*_{\pi})/2$; 5 -- average visual radius 
of NSNR in exponential medium. 
{\bf b}: 1 -- ratio of maximal to minimal diameters of visual shape of 
shock wave front; 2 -- ratio $R_{0}/R_{\pi}$; 3 -- $\log(\rho_{\pi}/\rho_0)$; 
4 -- $\log(T_{0}/T_{\pi}).$
          }
\label{evol-form-exp}
\end{figure*}
%%%%%%%%%%%%%%%%%%%%%%%%%%%%%%%%%%%%%%%%%%%%%%%%%%%%%%%%%%%%%%%%%%%%%%%%%%%%%%

\subsection{Shapes of SNRs in a flat exponential medium}

The exponential law distribution is frequently encountered in nature, 
especially in galactic disks. We consider the evolution of the shape 
of the shock front from a point explosion in a 
medium with exponential density distribution Eq.~(\ref{exp-density}). 
The morphological evolution of such SNR are presented in 
Fig.~\ref{shape-exp}. We can see from this figure the remarkable 
insensitivity of the {\em visible} form of the SNR to the ISM 
density gradient. \par

The apparent center of the non-spherical SNR does not coincide with 
real progenitor position. This may be important for localizing a 
possible compact stellar remnant (pulsar or black hole). \par

  The evolution of some shock characteristics is presented in 
Fig.~\ref{evol-form-exp}. The main result is that the average
visible morphological characteristics of the non-spherical SNRs 
are usually close to those for Sedov SNRs with the same initial 
parameters.  \par

%%%%=== Fig ===%%%%%%%%%%%%%%%%%%%%%%%%%%%%%%%%%%%%%%%%%%%%%%%%%%%%%%%%%%%%%%%
\begin{figure*}
%\picplace{9 cm}
\epsfysize=8.7truecm
\epsfbox{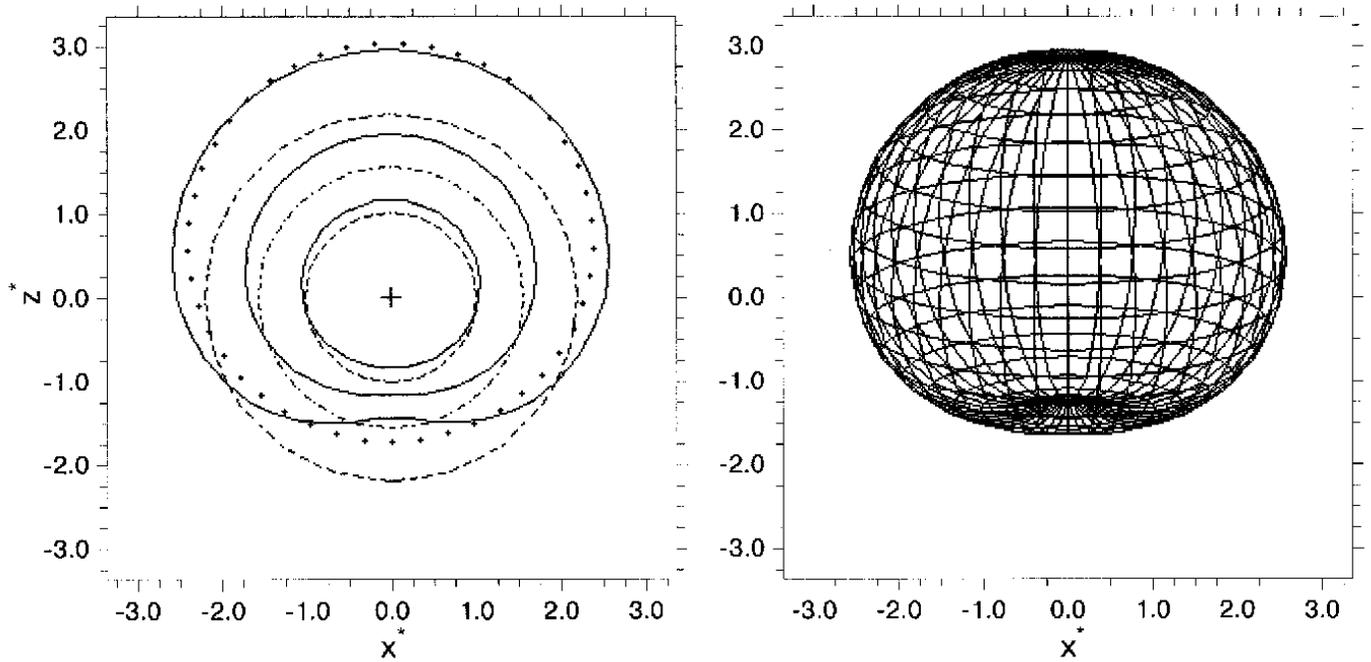}
\caption[]{
{\bf a and b}~The same as in Fig.~\ref{shape-exp} for a power 
law density distribution Eq.~(\ref{r^w-density}) for $r_0^*=1.75$ and $\tau=7.$ 
The angle between the symmetry axis of NSNR and the plane 
of the sky in the right figure equals $15^{\rm o}$. 
           }
\label{shape-r^w}
\end{figure*}
%%%%%%%%%%%%%%%%%%%%%%%%%%%%%%%%%%%%%%%%%%%%%%%%%%%%%%%%%%%%%%%%%%%%%%%%%%%%%%

\subsection{Shapes of SNRs in a power-law medium}

Another widely-used density distribution is the power-law one, created by
stellar winds, previous SN explosions etc.:
\begin{equation}
\label{r^w}
\rho^{\rm o}(\tilde r)=\rho_{\rm o}(\tilde r/R_{\rm m})^{\omega}.
\end{equation}
We consider the evolution of the shock from a point explosion in a medium 
with a spherically-symmetrical power-law density distribution when the 
explosion point is 
displaced by a distance $r_{\rm o}$ from the center of symmetry $\tilde r=0$ 
(wind source etc.). Therefore the density distribution as a function of 
the distance $r$ from the explosion point $r_{\rm o}$ is:
\begin{equation}
\label{r^w-density}
\rho^{\rm o}(r,\theta)=\rho^{\rm o}(0){\left(
{\sqrt{r_{\rm o}^2+r^2-2rr_{\rm o}\cos\theta}\over r_{\rm o}}
\right)}^w.
\end{equation}
Here $\rho^{\rm o}(0)$ is the initial density in the point of explosion.  
Hereafter, we take  $w=-2$ and $r_{\rm o}=1.75R_{\rm m}$. \par

The shape evolution of the SNR in this density distribution is shown in 
Fig.~\ref{shape-r^w}. It is worthy to note that  visible shapes of such
SNRs may be elongated transverse to the density gradient. \par

As one can see from Fig.~\ref{shape-r^w}, b like to the previous case 
(Fig.~\ref{shape-exp}, b), the projection of the SNR on the plane of the sky 
can cause a spherization of the visible SNR shape. 
So, even a visible spherical shape of SNR does not guarantee the uniformity 
of ISM and isotropy of explosion.\par

\subsection{Discussion}

From the above results, it follows that: \par
   1. The non-uniformity of the surrounding medium causes asphericity of SNRs.
The visible shapes may be elongated not only along (Fig.~\ref{shape-exp})
but also transverse to (Fig.~\ref{shape-r^w}) the density gradient, 
depending on the type of density distribution. \par

   2. SNR may have an apparent shape close to spherical even in cases of 
essential anisotropy of the real form and essential gradient of density 
distribution along the surface. \par

   3. The observed anisotropy of the shape is smaller than the real 
anisotropy as result of projection. The visible shape remains, generally, 
close to spherical. \par

   4. The non-uniformity of the surrounding medium results in differences 
of shock
characteristics along the SNR surface. If the initial density $\rho$ varies
along the shock surface, the maximal contrasts are expected in the X-ray
surface brightness
($\propto\rho^2$), they are smaller for the temperature distribution
($\propto\rho$) and minimal in shock and postshock gas velocities
($\propto\rho^{-1/5}$) (Figs.~\ref{shape-exp}--\ref{shape-r^w}).  \par

Therefore to determine the real conditions inside and around SNR, 
it is necessary to use additional information about SNR. We consider further 
the X-ray observations as an effective tool for SNR diagnostics. \par

%%%%%%%%%%%%%%%%%%%%%%%%%%%%%%%%%%%%%%%%%%%%%%%%%%%%%%%%%%%%%%%%%%%%%%%%%
%%%%======================Section III================================%%%%
%%%%%%%%%%%%%%%%%%%%%%%%%%%%%%%%%%%%%%%%%%%%%%%%%%%%%%%%%%%%%%%%%%%%%%%%%

\section{Integrated characteristics of the X-ray radiation from 
	non-spherical SNRs}

   SNRs are powerful sources of X-ray radiation, therefore X-ray observations
of SNR give unique information about physical conditions inside the remnants.
We calculate here the X-ray luminosities in the energy ranges
$\varepsilon=h\nu>0.1\ {\rm keV}$  ($L_{\rm x}^{>0.1}$) and 
$\varepsilon>4.5\ {\rm keV}$
($L_{\rm x}^{>4.5}$) as well as the spectral index $\alpha$ 
at $\varepsilon=5\ {\rm keV}.$  \par

%%%%=== Table I ===%%%%%%%%%%%%%%%%%%%%%%%%%%%%%%%%%%%%%%%%%%%%%%%%%%%%%%%%%%%
\begin{table*}
\caption[]{ Luminosity in ranges $\varepsilon>0.1\ {\rm keV}$ and 
$\varepsilon>4.5\ {\rm keV}$ (in parenthesis) of SNR expanding 
in an exponential medium Eq.~(\ref{exp-density}) with $H=10\ {\rm pc}$ 
(upper lines) in comparison with
Sedov SNR (lower lines). The SN explosion energy is $E_{\rm 51}=1$.  
          }
\begin{flushleft}
\begin{tabular}{ccccccc}
\hline
\multicolumn{1}{c}{$\log(t_{\rm yrs})$}&$\tau(n_{\rm H}^{\rm o})^{1/2}$&\multicolumn{5}{c}{$\log(n_{\rm H}^{\rm o})$} \\
\cline{3-7}
&&-1&-0.5&0&0.5&1\\
\hline
2.5 &0.056 &-             &34.97 (34.66) &35.53 (35.28) &36.12 (35.88) &36.74 (36.42)  \\
    &      &-             &34.96 (34.64) &35.52 (35.27) &36.11 (35.87) &36.73 (36.42)  \\
\noalign{\smallskip}
3.0 &0.176 &34.65 (34.39) &35.26 (34.93) &35.92 (35.42) &36.62 (35.83) &37.35 (36.12)  \\
    &      &34.61 (34.37) &35.23 (34.92) &35.89 (35.41) &36.60 (35.83) &37.34 (36.13)  \\
\noalign{\smallskip}
3.5 &0.555 &35.25 (34.31) &35.95 (34.61) &36.68 (34.79) &37.43 (34.83) &38.19 (34.80)  \\
    &      &35.10 (34.33) &35.84 (34.63) &36.60 (34.79) &37.38 (34.82) &38.16 (34.78)  \\
\noalign{\smallskip}
3.75&0.987 &35.72 (33.94) &36.41 (34.09) &37.13 (34.10) &37.86 (34.06) &38.57 (34.11)  \\
    &      &35.47 (33.98) &36.24 (34.07) &37.02 (34.05) &37.79 (34.03) &38.54 (34.10)  \\
\noalign{\smallskip}
4.0 &1.755 &36.23 (33.40) &36.89 (33.41) &37.56 (33.37) &38.23 (33.43) &-        \\
    &      &35.88 (33.32) &36.66 (33.28) &37.42 (33.31) &38.15 (33.40) &-        \\
\noalign{\smallskip}
4.25&3.121 &36.74 (32.86) &37.30 (32.78) &37.88 (32.79) &-        &-         \\
    &      &36.29 (32.53) &37.04 (32.60) &37.74 (32.72) &-        &-         \\
\noalign{\smallskip}
4.5 &5.555&37.04 (32.54) &37.53 (32.40) &-        &-        &-        \\
    &     &36.65 (31.90) &37.33 (32.04) &-        &-        &-        \\
\hline
\end{tabular}
\end{flushleft}
\label{tab-a}
\end{table*}
%%%%%%%%%%%%%%%%%%%%%%%%%%%%%%%%%%%%%%%%%%%%%%%%%%%%%%%%%%%%%%%%%%%%%%%%%%%%%%

\subsection{Plasma X-ray emissivity under ionization equilibrium condition}

We assume the cosmic abundance (Allen \cite{Allen73}). 
The X-ray continuum energy emissivity per unit energy interval is  
(in ${\rm erg\ cm^{-3}\ s^{-1}\ keV^{-1}}$) 
\begin{equation} 
\label{cont-emiss} 
\begin{array}{l}
P_{\rm c}(T,\varepsilon)=  \\ \\
{\displaystyle
\qquad =1.652\cdot 10^{-23}n_{\rm e}^2 G_{\rm c}(T,\varepsilon)
           T_6^{-1/2}exp\left({-11.59\varepsilon \over T_6} \right),
}
\end{array}
\end{equation}
where $\varepsilon>0.1$ is the photon energy in {\rm keV}, $T_6$ is the 
plasma temperature
in $10^6\ {\rm K}$, $n_{\rm e}$ is the electron number density. 
The approximation for total Gaunt factor $G_{\rm c}$ as the sum 
of Gaunt factors for free-free $G_{\rm ff},$ free-bound $G_{\rm fb}$ 
and two-photon $G_{\rm 2\gamma}$ processes was taken from
Mewe et al.  (\cite{Mewe-Lem-Oord86}):  
\begin{equation} 
\label{Gaunt-factor}
G_{\rm c}(T,\varepsilon)=27.83(T_6+0.65)^{-1.33}+
        0.35{\varepsilon }^{-0.34}T_6^{0.422}.
\end{equation}
This approximation represents the continuum losses to an 
accuracy of $10 - 20\%$ for $T\ge 3\cdot 10^6\ {\rm K}$ and of $30 - 50\%$ 
for $T=(0.2 - 3)\cdot 10^6\ {\rm K}$. \par

   For calculation of continuum and line emission of plasma in the
different energy ranges, we have approximated the Raymond \& Smith
(\cite{Raym-Smith77}) data for plasma emissivity $\Lambda$
(in ${\rm erg\ cm^{-3}\ s^{-1}}$) \newline
for $\varepsilon=0.1 - 2.4\ {\rm keV}$ as follows: 
\begin{equation}
\label{approx-0.1-2.4}
\Lambda^{0.1-2.4}(T)=10^{-21.80}T_6^{-0.63}\exp\left( {-1.40 \over T_6}
\right),
\end{equation}
(accurate to $5 - 30\%$ for $\log T=5.3 - 8$);\newline
for $\varepsilon>4.5\ {\rm keV}$ as follows: 
\begin{equation}
\label{approx->4.5}
\Lambda^{>4.5}(T)=10^{-22.42}\exp\left( {-39.50 \over T_6^{0.77}}
\right),
\end{equation}
(accurate to $1 - 8\%$ for $\log T=6.7 - 8$); \newline
for $\varepsilon>0.1\ {\rm keV}$ as follows: 
\begin{equation}
\label{approx->0.1}
\Lambda^{>0.1}(T)=\Lambda^{0.1-2.4}(T)+10^{-24.4}T_6^{0.8}
,
\end{equation}
(accurate to $7 - 35\%$ for $\log T=5.5 - 8$). \par

The total flux $F_{\varepsilon}$ at photon energy $\varepsilon$ and 
luminosity $L_{\rm x}$ of the entire SNR can be calculated by integrating 
over the remnant volume $V$ 
\begin{equation}
\label{contin_flux}
F_{\varepsilon}=\int\limits_V
P_{\rm c}(T,\varepsilon) dV ,
\end{equation}
\begin{equation}
\label{luminosity}
L_{\rm x}=\int\limits_V
\Lambda(T) n_{\rm e} n_{\rm H} dV .
\end{equation}
The spectral index $\alpha$ is 
\begin{equation}
\label{alpha-def}
\alpha=-{\partial \ln F_{\varepsilon} \over \partial \ln \varepsilon}.
\end{equation}

\subsection{Evolution of the total X-ray emission from aspherical SNRs}

  It is well-known that the Sedov (\cite{Sedov}) solution for SNR 
characteristics in a uniform medium is determined by three parameters, 
e.g., the energy of the explosion 
$E_{\rm o}=10^{51}E_{51}\ {\rm erg},$ the initial number density 
$n_{\rm H}^{\rm o}$ and time $t$. 
The shape of the integrated spectrum (in particular the spectral 
index $\alpha$) depends only on one 
parameter (the shock temperature $T_{\rm s}$) in case of 
collision ionization equilibrium (CIE) and on two 
($T_{\rm s}$ and $\eta=E_{\rm o}\left(n_{\rm H}^{\rm o}\right)^2$) 
in case of NEI (for fixed abundance) (Hamilton et al. 
\cite{Hamilton}). In order to estimate the total luminosity of 
a Sedov SNR in the NEI case, we need a third parameter, i.e. the explosion 
energy $E_{\rm o}$ (Hamilton et al. \cite{Hamilton}). In the Sedov CIE case 
the luminosity depends only on two parameters 
$T_{\rm s}$ and $\zeta=E_{\rm 51}n_{\rm H}^{\rm o}.$ \par
 
  If the SN explodes in a non-uniform medium, we have an additional 
fourth parameter which characterises the non-uniform density distribution. 
In our cases, it is the scale height $H$ for exponential density 
distribution or $r_{\rm o}$ for power-law density distribution. \par

%%%%=== Table II ===%%%%%%%%%%%%%%%%%%%%%%%%%%%%%%%%%%%%%%%%%%%%%%%%%%%%%%%%%%
\begin{table}
\caption[]{ Spectral index $\alpha(5\ keV)$ of NSNR in exponential medium
Eq.~(\ref{exp-density}) with $H=10\ pc$ (upper lines) in comparison with
Sedov SNR (lower lines). The SN explosion energy is $E_{\rm 51}=1$.  }
\begin{flushleft}
\begin{tabular}{ccccccc}
\hline
\multicolumn{1}{c}{$\log(t_{\rm yrs})$}&$\tau(n_{\rm H}^{\rm o})^{1/2}$&\multicolumn{5}{c}{$\log(n_{\rm H}^{\rm o})$}\\
\cline{3-7}
&&-1&-0.5&0&0.5&1\\
\hline
2.5 &0.056 &- &0.49 &0.58 &0.71 &0.92  \\
    &      &- &0.49 &0.57 &0.70 &0.91  \\
\noalign{\smallskip}
3.0 &0.176 &0.75 &0.95 &1.26 &1.74 &2.47  \\
    &      &0.70 &0.91 &1.23 &1.72 &2.45  \\
\noalign{\smallskip}
3.5 &0.555 &1.95 &2.57 &3.46 &4.52 &5.05  \\
    &      &1.72 &2.45 &3.49 &4.67 &5.06  \\
\noalign{\smallskip}
3.75 &0.987 &3.06 &3.81 &4.70 &4.89 &4.24  \\
     &      &2.93 &4.10 &5.04 &4.78 &4.22  \\
\noalign{\smallskip}
4.0 &1.755 &3.80 &4.56 &4.73 &4.07 &-  \\
    &      &4.67 &5.06 &4.45 &4.05 &-  \\
\noalign{\smallskip}
4.25 &3.121 &4.02 &4.55 &4.01 &- &-  \\
     &      &4.78 &4.21 &3.94 &- &-  \\
\noalign{\smallskip}
4.5 &5.555 &3.60 &3.82 &- &- &-  \\
    &      &4.05 &3.85 &- &- &-  \\
\hline
\end{tabular}
\end{flushleft}
\label{tab-b}
\end{table}
%%%%%%%%%%%%%%%%%%%%%%%%%%%%%%%%%%%%%%%%%%%%%%%%%%%%%%%%%%%%%%%%%%%%%%%%%%%%%
%%%%=== Table III ===%%%%%%%%%%%%%%%%%%%%%%%%%%%%%%%%%%%%%%%%%%%%%%%%%%%%%%%%%
\begin{table}
\caption[]{ The same as in Table~\ref{tab-a} for $E_{\rm 51}=0.1$ and
range $\varepsilon>0.1\ {\rm keV}$.  
          } 
\begin{flushleft}
\begin{tabular}{ccccccc}
\hline
\multicolumn{1}{c}{$\log(t_{\rm yrs})$}&$\tau(n_{\rm H}^{\rm o})^{1/2}$&\multicolumn{5}{c}{$\log(n_{\rm H}^{\rm o})$}\\
\cline{3-7}
&&-1&-0.5&0&0.5&1\\
\hline
2.5 & 0.018&- &34.12 &34.74 &35.40 &36.10  \\
    &      &- &34.11 &34.73 &35.39 &36.10  \\
\noalign{\smallskip}
3.0 &0.056 &33.92 &34.62 &35.35 &36.11 &36.89  \\
    &      &33.89 &34.60 &35.34 &36.11 &36.88  \\
\noalign{\smallskip}
3.5 &0.176 &34.68 &35.43 &36.19 &36.94 &37.66  \\
    &      &34.60 &35.38 &36.16 &36.92 &37.65  \\
\noalign{\smallskip}
4.0 &0.555 &35.56 &36.23 &36.86 &37.43 &37.90  \\
    &      &35.42 &36.15 &36.83 &37.43 &37.90  \\
\hline
\end{tabular}
\end{flushleft}
\label{tab-c}
\end{table}
%%%%%%%%%%%%%%%%%%%%%%%%%%%%%%%%%%%%%%%%%%%%%%%%%%%%%%%%%%%%%%%%%%%%%%%%%%%%%%
%%%%=== Table IV ===%%%%%%%%%%%%%%%%%%%%%%%%%%%%%%%%%%%%%%%%%%%%%%%%%%%%%%%%%%
\begin{table}
\caption[]{ The same as in Table~\ref{tab-b} for $E_{\rm 51}=0.1$.
           }
\begin{flushleft}
\begin{tabular}{ccccccc}
\hline
\multicolumn{1}{c}{$\log(t_{\rm yrs})$}&$\tau(n_{\rm H}^{\rm o})^{1/2}$&\multicolumn{5}{c}{$\log(n_{\rm H}^{\rm o})$}\\
\cline{3-7}
&&-1&-0.5&0&0.5&1\\
\hline
2.5 &0.018 &- &0.71 &0.92 &1.23 &1.72  \\
    &      &- &0.70 &0.91 &1.23 &1.72  \\
\noalign{\smallskip}
3.0 &0.056 &1.26 &1.74 &2.47 &3.49 &4.66  \\
    &      &1.23 &1.72 &2.45 &3.49 &4.67  \\
\noalign{\smallskip}
3.5 &0.176 &3.46 &4.52 &5.05 &4.48 &4.05  \\
    &      &3.49 &4.67 &5.06 &4.45 &4.05  \\
\noalign{\smallskip}
4.0 &0.555 &4.73 &4.07 &3.87 &3.76 &3.68  \\
    &      &4.45 &4.05 &3.85 &3.74 &3.66  \\
\hline
\end{tabular}
\end{flushleft}
\label{tab-d}
\end{table}
%%%%%%%%%%%%%%%%%%%%%%%%%%%%%%%%%%%%%%%%%%%%%%%%%%%%%%%%%%%%%%%%%%%%%%%%%
%%%%=== Table V ===%%%%%%%%%%%%%%%%%%%%%%%%%%%%%%%%%%%%%%%%%%%%%%%%%%%%%%%%%%%
\begin{table}[t]
\caption[]{
The ratios of SNR characteristics (the volume $V$, swept-up mass 
$M$ and characteristic temperatures $T_{\rm ch}$)
in non-uniform medium (with exponential (E) Eq.~(\ref{exp-density}) 
and power law (PL) Eq.~(\ref{r^w-density}) density distribution) 
to uniform ones.
	}
\begin{flushleft} 
\begin{tabular}{ccccccccc} \hline
$\tau$&\multicolumn{2}{c}{$V/V_{{\rm s}}$}&&\multicolumn{2}{c}{$M/M_{{\rm s}}$}&&\multicolumn{2}{c}{$T_{\rm ch}/T_{s}$} \\
\cline{2-3}\cline{5-6}\cline{8-9}
&E&PL&&E&PL&&E&PL\\
\cline{2-3}\cline{5-6}\cline{8-9}
\hline
0.1  &1.00&1.02 &&1.00&1.00 &&1.00&1.00  \\
0.5  &1.03&1.06 &&1.00&0.99 &&1.00&1.01  \\
1.0  &1.05&1.10 &&1.01&0.98 &&0.99&1.02  \\
3.0  &1.14&1.24 &&1.01&0.95 &&0.99&1.06  \\
5.0  &1.23&1.37 &&1.02&0.92 &&0.98&1.09  \\
10.0 &1.51&1.64 &&1.04&0.86 &&0.96&1.16  \\
20.0 &2.53&2.13 &&1.07&0.75 &&0.94&1.33  \\
\hline
\end{tabular}
\end{flushleft}
\label{tab-e}
\end{table}
%%%%%%%%%%%%%%%%%%%%%%%%%%%%%%%%%%%%%%%%%%%%%%%%%%%%%%%%%%%%%%%%%%%%%%%%%

   It is naturally to expect, that in cases of NSNR considered, 
the total X-ray luminosity $L_{\rm x}$ as well as spectral index
$\alpha$ will strongly depend on non-uniformity characteristics, such as
scale height $H$ etc. Therefore, we calculate here extensive grid of 
models for
evolution of X-ray radiation from NSNRs which evolve in non-uniform ISM and
compare it with the Sedov case.  Our grid covers the range of paramers 
which correspond to the adiabatic stage of SNR evolution, 
i.e. for SNR radii between 
$R_{\rm i}=\left(3M_{\rm ej}/4\pi\rho^{\rm o}(0)\right)^{1/3}
\simeq2.5(n_{\rm H}^{\rm o})^{-1/3}\ {\rm pc}$ when the swept up mass of ISM 
becomes equal to the mass of ejecta $M_{\rm ej}\simeq1M_{\sun},$ and 
$R_{\rm f}$ when the effective temperature $T_{\rm ef}$ 
corresponds to the maximum value of plasma emissivity function 
$T_{\rm ef}\simeq1.3T_{\rm s}=5.2\cdot 10^{5}\ {\rm K}$ 
(Lozinskaya \cite{Lozins}). 
In a non-uniform medium, the value of $R_{\rm f}$ depends on the concrete 
density distribution in a selected direction (sector). 
So, if the density distribution in a sector is exponential 
$n_{\rm H}^{\rm o}(r)=n_{\rm H}^{\rm o}(0)\cdot\exp(\pm r/H),$ 
we may estimate 
$R_{\rm f}$ from the relation $R_{\rm f}\cdot \exp(\pm R_{\rm f}/3H)=
25(E_{\rm 51}/n_{\rm H}^{\rm o}(0))^{1/3}\ {\rm pc}.$ 
For a uniform medium $H=\infty,$ we obtain  
$R_{\rm max}\simeq 25E_{\rm 51}^{0.3}(n_{\rm H}^{\rm o}(0))^{-0.3}\ {\rm pc},$ 
which is close to results of other authors (Lozinskaya \cite{Lozins}).
In our calculation, the maximal time was estimated from adiabaticity 
violation 
in the $\theta=\pi/2$ sector. For parameters of ISM considered here, 
maximal radii of SNRs do not exceed a few scale heights. 
Therefore, obtained results are not 
affected by shock acceleration, which is important only at considerably 
greater distances. \par

%%%%=== Fig ===%%%%%%%%%%%%%%%%%%%%%%%%%%%%%%%%%%%%%%%%%%%%%%%%%%%%%%%%%%%%%%%
\begin{figure*}[t]
%\picplace{8 cm}
\epsfysize=8truecm
\epsfbox{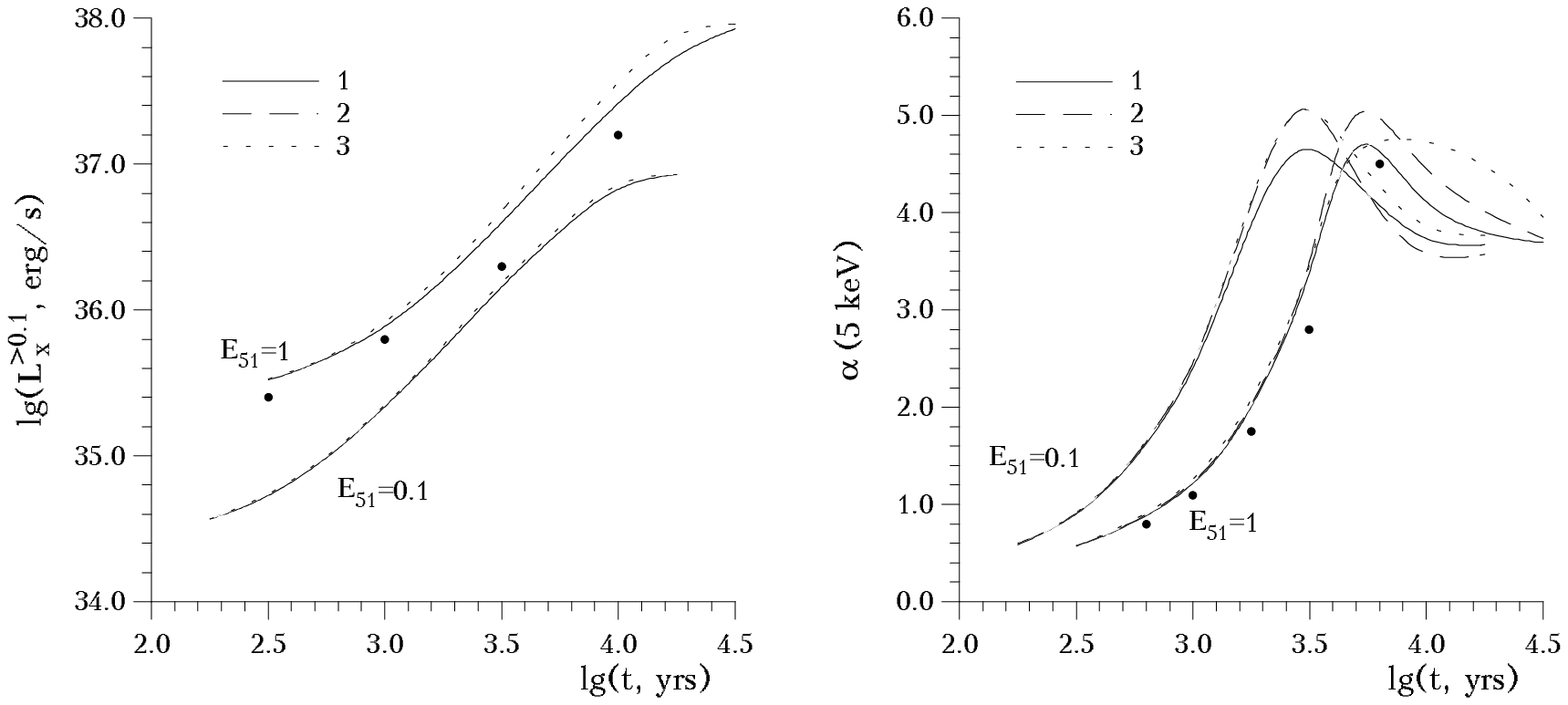}
\caption[]{
Evolution of luminosity $L_{\rm x}$ 
in range $>0.1\ {\rm keV}$ and of 
spectral index $\alpha(5\ {\rm keV})$ of SNRs in a uniform
medium (line 1), in an exponential medium Eq.~(\ref{exp-density}) with
$H=80\ {\rm pc}$ (line 2) and $H=10\ {\rm pc}$ (line 3). In all 
cases $n_{\rm H}^{\rm o}(0)=1\ {\rm cm^{-3}}$. 
Results of numerical calculations of Hamilton et al. (\cite{Hamilton}) 
for a uniform medium and $E_{51}=1$ with Meyer abundance are also 
shown by dots. Hereafter $\gamma=5/3$.  
          } 
\label{lum-alpha-evol}
\end{figure*}
%%%%%%%%%%%%%%%%%%%%%%%%%%%%%%%%%%%%%%%%%%%%%%%%%%%%%%%%%%%%%%%%%%%%%%%%%%%%%%

We show in Fig.~\ref{lum-alpha-evol} the evolution of X-ray luminosity
for range $\varepsilon>0.1\ {\rm keV}$ and spectral index $\alpha$ at 
$5\ {\rm keV}$
for SNR in an exponential medium. It is easy to see that both X-ray
characteristics of NSNR evolve analogously to the Sedov case and
similarity increases with decreasing surrounding medium
density gradient (with increasing $H$) and decreasing explosion energy. 
The results presented in Fig.~\ref{lum-alpha-evol} of 
calculations of the Sedov case for $E_{\rm 51}=1$ 
(Hamilton et al. \cite{Hamilton}) reveal 
some differences between X-ray luminosities at low temperatures. 
They may be caused, in part, by atomic data and chemical composition 
differences (our model uses the Allen abundance, whereas in the paper 
of Hamilton et al. (\cite{Hamilton})
have been used the Meyer abundance). Our results for Sedov case 
coinside with of Leahy \& Aschenbach (\cite{Leahy-Asch-96}) and 
Kassim et al. (\cite{Kessim-Hertz-Weiler-93}) results for 
model with Allen abundance within 25\%. \par

%%%%=== Table VI ===%%%%%%%%%%%%%%%%%%%%%%%%%%%%%%%%%%%%%%%%%%%%%%%%%%%%%%%%%%
{\small
\begin{table*}
\caption[]{
Evolution of SNR luminosity $L_{\rm x}$ and spectral 
index $\alpha(5\ {\rm keV})$ 
in uniform (S) and  exponential (E) Eq.~(\ref{exp-density}) media.}
\begin{flushleft}
\begin{tabular}{ccccccccccccccc}
\hline
&&&&\multicolumn{11}{c}{$\log\zeta=\log(n_{\rm H}^{\rm o}\cdot E_{\rm 51})$}\\
\cline{5-15}
$\log(T_{\rm ch})$&model&H&$E_{\rm 51}$&&-1&&&&0&&&&1&\\
\cline{5-7}\cline{9-11}\cline{13-15}
&&&&$\tau$&$L_{\rm x}^{>0.1}$&$\alpha_5$&&$\tau$&$L_{\rm x}^{>0.1}$&$\alpha_5$&&$\tau$&$L_{\rm x}^{>0.1}$&$\alpha_5$\\
\hline
8.0 &S &   &        &     &34.68&0.81 &&     &35.68&0.81 &&     &36.68&0.81  \\
    &E &80 &1       &0.004&34.68&0.82 &&0.001&35.68&0.82 &&8e-5 &36.68&0.81  \\
    &E &10 &0.1     &0.015&34.68&0.82 &&0.002&35.68&0.82 &&3e-4 &36.68&0.82  \\
    &E &10 &1       &0.696&34.73&0.88 &&0.102&35.69&0.83 &&0.015&36.68&0.82  \\
\noalign{\smallskip}
7.2 &S &   &        &     &35.49&3.02 &&     &36.49&3.02 &&     &37.49&3.02  \\
    &E &80 &1       &0.018&35.50&3.03 &&0.003&36.50&3.03 &&4e-4 &37.49&3.02  \\
    &E &10 &0.1     &0.070&35.51&3.03 &&0.010&36.50&3.03 &&0.001&37.50&3.03  \\
    &E &10 &1       &3.232&35.75&3.13 &&0.474&36.56&3.05 &&0.070&37.51&3.03  \\
\noalign{\smallskip}
6.4 &S &   &        &     &36.56&4.18 &&     &37.56&4.18 &&      &38.56&4.18  \\
    &E &80 &1       &0.083&36.57&4.19 &&0.012&37.56&4.18 &&0.002&38.56&4.18  \\
    &E &10 &0.1     &0.323&36.59&4.20 &&0.047&37.57&4.18 &&0.007&38.56&4.18  \\
    &E &10 &1       &15.00&36.98&3.73 &&2.202&37.70&4.38 &&0.323&38.59&4.20  \\
\hline
\end{tabular}
\end{flushleft}
\label{tab-f}
\end{table*}
}
%%%%%%%%%%%%%%%%%%%%%%%%%%%%%%%%%%%%%%%%%%%%%%%%%%%%%%%%%%%%%%%%%%%%%%%%%

We have calculated a grid of NSNR models (Tables~\ref{tab-a}-\ref{tab-d})
which confirms that this analogy in evolution is an intrinsic property of 
NSNR X-ray 
radiation. In fact, it may be seen from Tables~\ref{tab-a}-\ref{tab-d}
that for a wide range of number density at the point of explosion
$n_{\rm H}^{\rm o}(0)=0.1 - 10\ {\rm cm^{-3}},$ the X-ray luminosity of 
adiabatic NSNR
evolving in a medium with strong enough density gradient ($H=10\ {\rm pc}$) 
in different energy ranges are not far from the X-ray luminosity of SNR in
a uniform density medium with the same initial model parameters. The
differences increase with age of the SNR and with decreasing 
$n_{\rm H}^{\rm o}(0)$. But
even for old NSNR (e.g., $t=3\cdot 10^{4}\ {\rm years}$) in a low density 
medium
($n_{\rm H}^{\rm o}(0)=0.1\ {\rm cm^{-3}}$), the maximal difference is 
about $60\% $ (for
$\tau=17.5$).  The differences in spectral index of adiabatic NSNR
do not exceed $10\% $.  Table~\ref{tab-a} shows that luminosity in range
$\varepsilon>4.5\ {\rm keV}$ is close to the Sedov case also.  These last 
two facts reveal the similarity of total spectra of Sedov SNR and
non-spherical ones. \par

We suppose that the behevior of integral X-ray characteristics of NSNR in
other cases of smooth continuous density distribution of surrounding medium 
will be similar considered above, since X-ray emission mainly 
depends on the emission measure $EM\simeq n_{\rm e}^2V\simeq M^2V^{-1},$ 
but both masses of swepted up gas $M$ and
volume of NSNR $V$ remain close to those of Sedov SNR (Table~\ref{tab-e}).
Moreover, the last fact allow us to introduce a characteristic
shock temperature for NSNR $T_{\rm ch}$, comparable to $T_{\rm s}$ in the 
Sedov case:
\begin{equation} 
\label{def-T_ch} 
T_{\rm ch}^*={32\pi \over 75}{{\gamma-1}\over {(\gamma+1)^2}}(M^*)^{-1},
\end{equation}
where
$T^*=T/T_{\rm m},$ $M^*=M/M_{\rm m}$ with
$T_{\rm m}=\alpha_{\rm A}(2,\gamma)^{-1}A_{\rm gas}^{-1}E_{\rm o}\rho_{\rm o}^{-1}(0)R_{\rm m}^{-3}$ and
$M_{\rm m}=\rho_{\rm o}(0)R_{\rm m}^3.$
The temperature $T_{ef}$ determined from the observed spectrum will be
connected with $T_{\rm ch}$ as in the Sedov case $T_{ef}\approx 1.3 T_{\rm ch}$
(Itoh \cite{Itoh77}). This characteristic temperature may be used to
estimate the parameters of the whole remnant, such as mass, volume, etc. \par

 The results of calculations show that integral X-ray characteristics 
weakly depend on the surrounding medium density gradient (e.g., $H$).
Therefore, as in the Sedov case, a spectral index of equilibrium emission 
from NSNR depends approximately only on one parameter $T_{\rm ch}$ and luminosity 
depends approximately only on two parameters: 
$T_{\rm ch}$ and $\zeta=E_{\rm 51}n_{\rm H}^{\rm o}(0)$. 
As confirmation of this,
we show that maximal deviation from this rule for NSNR in exponential
density distribution reaches few percent for spectral index and a 
few tens of percent for luminosity (Table~\ref{tab-f}).  \par

\subsection{Discussion}

From Fig.~\ref{lum-alpha-evol} and
Tables~\ref{tab-a}-\ref{tab-f} follows a remarkable fact of proximity of
total fluxes and spectrum shapes of SNRs in uniform and non-uniform
media even in case of considerable anisotropy of adiabatic NSNR. The
reason of this phenomenon is mutual compensation of emission deficit from
low density regions of NSNR and enhanced emission from high density 
regions. This explains the strange circumstance that integral X-ray
characteristics of the majority of SNRs with evident asymetry in shape 
and/or surface brightness distribution are
described with sufficient accuracy by the Sedov model of SN explosion in
a uniform medium. For more correct modelling of physical conditions inside 
and outside the SNR, it is necessary to analyze the characteristics of X-ray 
emission distributed over the SNR surface. \par

%%%%%%%%%%%%%%%%%%%%%%%%%%%%%%%%%%%%%%%%%%%%%%%%%%%%%%%%%%%%%%%%%%%%%%%%%
%%%%======================Section IV=================================%%%%
%%%%%%%%%%%%%%%%%%%%%%%%%%%%%%%%%%%%%%%%%%%%%%%%%%%%%%%%%%%%%%%%%%%%%%%%%

%%%%=== Fig ===%%%%%%%%%%%%%%%%%%%%%%%%%%%%%%%%%%%%%%%%%%%%%%%%%%%%%%%%%%%%%%%
\begin{figure}
%\picplace{9 cm}
\epsfysize=8truecm
\epsfbox{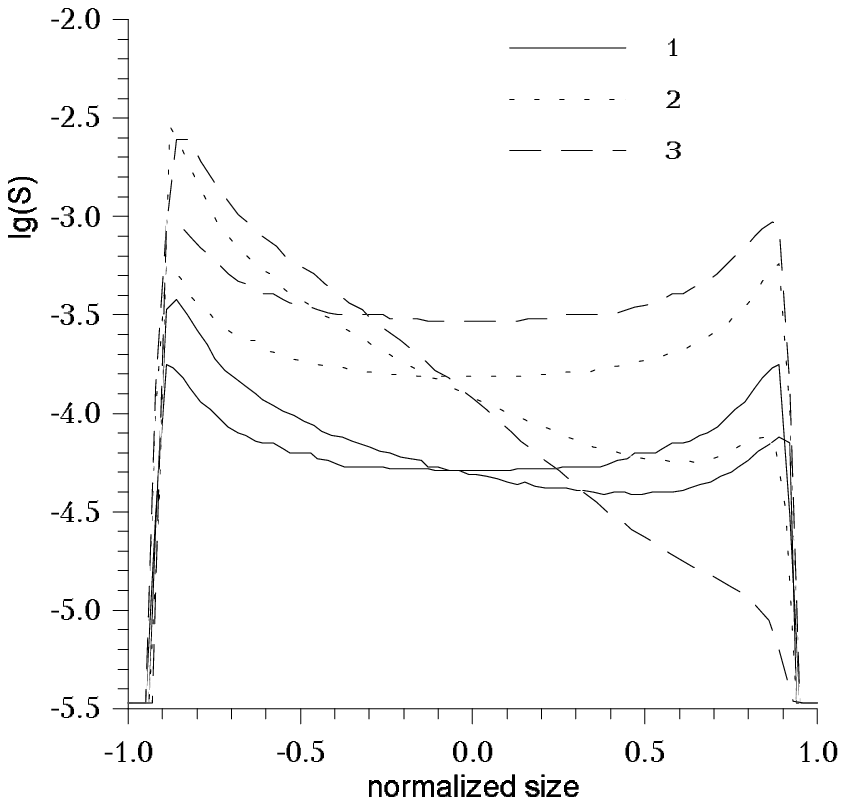}
\caption[]{
Distribution of $\varepsilon\!>\!0.1\ {\rm keV}$ surface brightness 
$S^{>0.1}$ along the NSNR symmetry axis  in uniform and exponential 
Eq.~(\ref{exp-density}) media 
for time moments $t=1000$ {\rm yrs} (line 1), 5000 {\rm yrs} (line 2), 
40000 {\rm yrs} (line 3). The model parameters are $E_{\rm 51}=1,$ 
$n_{\rm H}^{\rm o}(0)=1\ {\rm cm^{-3}},$ $H=10\ {\rm pc}.$
           }
\label{surf-brigh-a}
\end{figure}
%%%%%%%%%%%%%%%%%%%%%%%%%%%%%%%%%%%%%%%%%%%%%%%%%%%%%%%%%%%%%%%%%%%%%%%%%%%%%%
%%%%=== Fig ===%%%%%%%%%%%%%%%%%%%%%%%%%%%%%%%%%%%%%%%%%%%%%%%%%%%%%%%%%%%%%%%
\begin{figure}
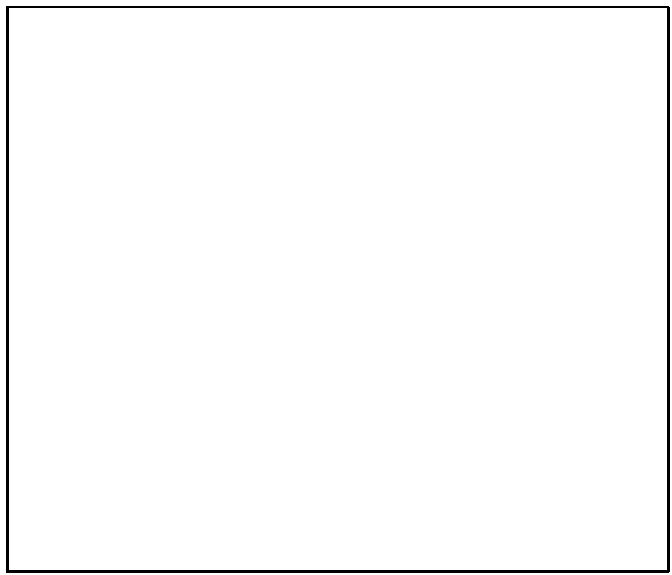

\picplace{7.5 cm}
%\epsfysize=7.5truecm
%\epsfbox{pcs_9.eps}
\caption[]{
Distribution of surface brightness $S^{>0.1}$ of the Sedov SNR in 
uniform medium in the range $\varepsilon>0.1\ {\rm keV}$. 
The model parameters are $E_{\rm 51}=1,\ n_{\rm H}^{\rm o}(0)=0.2\
{\rm cm^{-3}},\ t=1000\ {\rm yrs}$.  The SNR characteristics are 
$\log (L_{\rm x}^{>0.1},\ {\rm erg/s})=34.98,$ 
$M=9.4\ {\rm M_\odot},$ $T_{\rm s} =9.9\cdot 10^7\ {\rm K}$.  
The lines of constant brightness are indicated by values of 
logarithm of flux $\log(S)$. 
The center of explosion hereafter is at the origin of the coordinates. 
          }
\label{surf-map}
\end{figure}
%%%%%%%%%%%%%%%%%%%%%%%%%%%%%%%%%%%%%%%%%%%%%%%%%%%%%%%%%%%%%%%%%%%%%%%%%%%%%%
%%%%=== Fig ===%%%%%%%%%%%%%%%%%%%%%%%%%%%%%%%%%%%%%%%%%%%%%%%%%%%%%%%%%%%%%%%
\begin{figure}
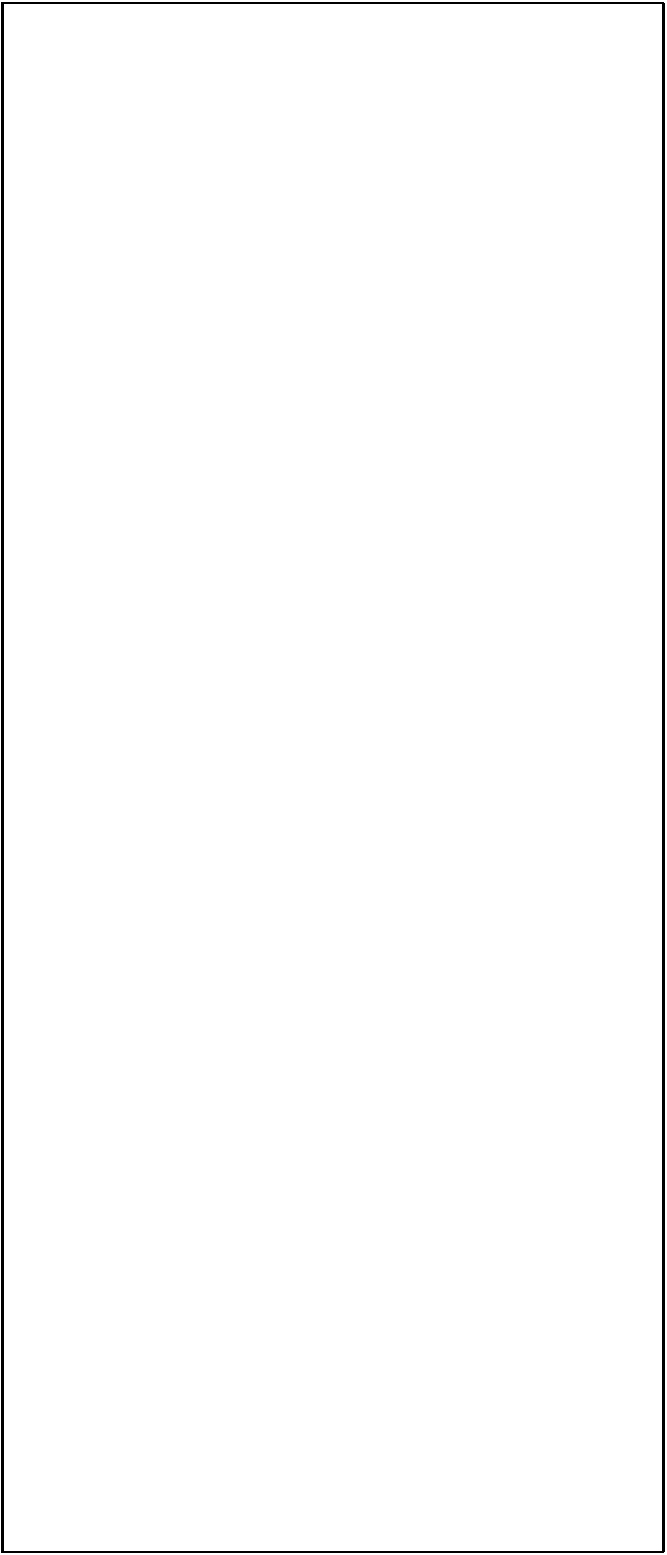

\picplace{20.6 cm}
%\epsfysize=20.6truecm
%\epsfbox{pc_10.eps}
\caption[]{
The same as in Fig.~\ref{surf-map} for exponential medium 
Eq. (\ref{exp-density}) 
with $H=10\ {\rm pc}$. The NSNR parameters at this time  
are $\log(L_{\rm x}^{>0.1},\ {\rm erg/s})=35.02,\
M=9.4\ {\rm M_\odot},\ T_{\rm ch}=9.9\cdot 10^7\ {\rm K}$.  
The NSNR inclination angle 
to the plane of the sky equals $0^{\rm o}$ (upper case), $45^{\rm o}$ 
(center case) and $90^{\rm o}$ (lower case).
          }
\label{surf-map-a}
\end{figure}
%%%%%%%%%%%%%%%%%%%%%%%%%%%%%%%%%%%%%%%%%%%%%%%%%%%%%%%%%%%%%%%%%%%%%%%%%%%%%%   
\clearpage
%%%%=== Fig ===%%%%%%%%%%%%%%%%%%%%%%%%%%%%%%%%%%%%%%%%%%%%%%%%%%%%%%%%%%%%%%%
\begin{figure*}
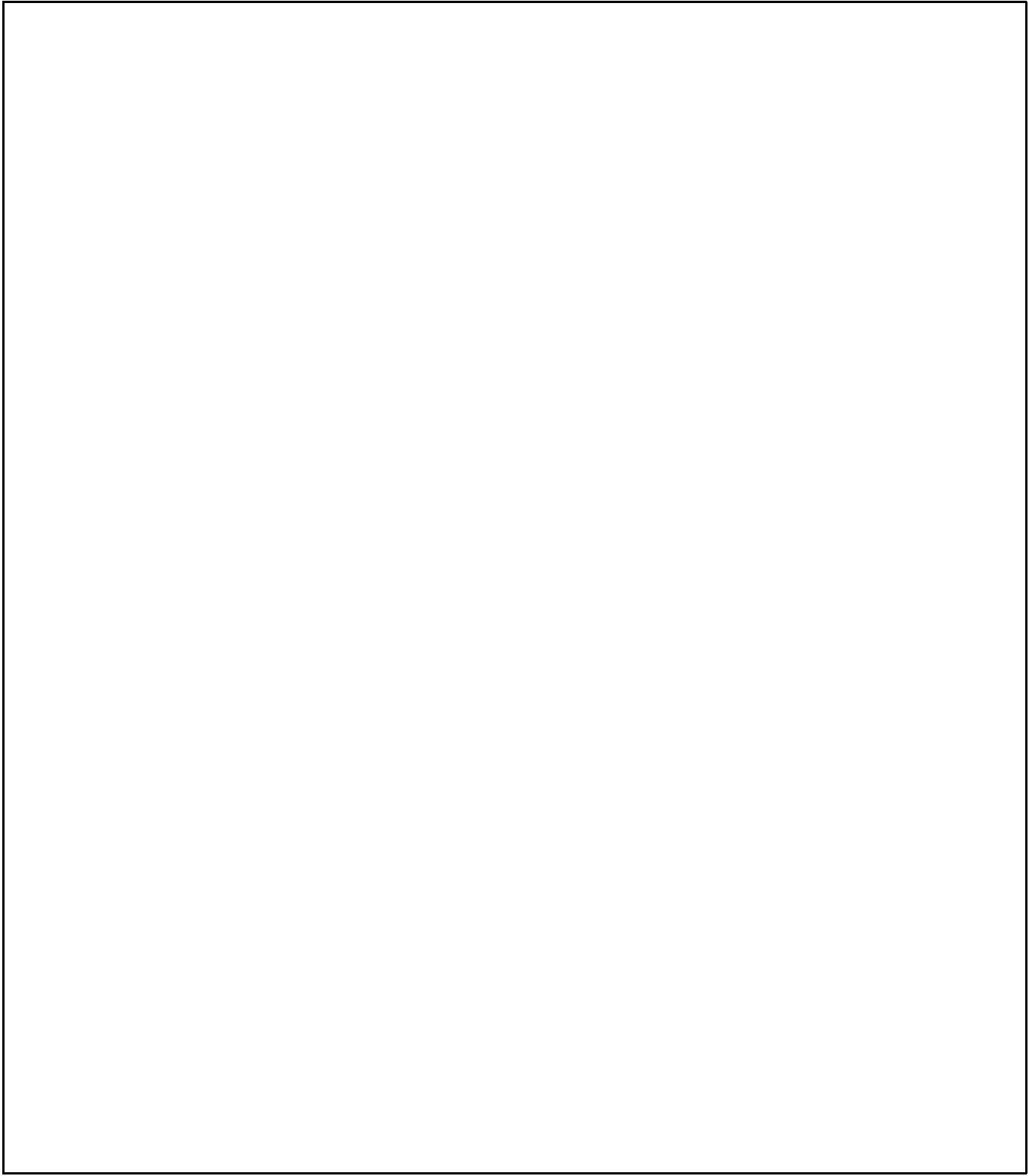

\picplace{20.60 cm}
%\epsfysize=20.6truecm
%\epsfbox{pc_11.eps}
\caption[]{
The same as in Fig.~\ref{surf-map-a} for time $t=3300\ {\rm yrs}$
($\log(L_{\rm x}^{>0.1})=35.70,\ M=40.0\ {\rm M_\odot},\ T_{\rm ch}=2.3\cdot 10^7\ {\rm K}$) and
$t=25500\ {\rm yrs}$ ($\log(L_{\rm x}^{>0.1})=37.28,\ M=476\ {\rm M_\odot},\ T_{\rm s}=2.0\cdot
10^6\ {\rm K}$).
          }
\label{surf-map-bc}
\end{figure*}
%%%%%%%%%%%%%%%%%%%%%%%%%%%%%%%%%%%%%%%%%%%%%%%%%%%%%%%%%%%%%%%%%%%%%%%%%%%%%%
\clearpage

\section{Surface - distributed characteristics of NSNR X-ray emission}

\subsection{Surface brightness in range $\varepsilon>0.1\ {\rm keV}$} \par

We analyse here the surface brightness distribution $S(x,z)$ 
(in ${\rm erg\ s^{-1}\ cm^{-2}\ ster^{-1}}$)

\begin{equation}
S(x,z)={1\over 4\pi}\int\limits_{y_1}^{y_2}\Lambda(T) n_{\rm e} n_{\rm H} dy
\label{surf-bright}
\end{equation}
and surface distribution of spectral index 
\begin{equation}
\alpha(x,z)=-{\partial \ln \over \partial \ln \varepsilon}
            \int\limits_{y_1}^{y_2} P_{\rm c}(T,\varepsilon) dy, 
\label{surf-alpha}
\end{equation}
where the integrals are taken along the line of sight inside the remnant.

\subsubsection{NSNRs in a medium with flat exponential density distribution} 

  Figs.~\ref{surf-brigh-a}--\ref{surf-map-bc}
demonstrate an evolution of surface brightness of NSNR 
in media with exponential density distribution (\ref{exp-density}).
At the beginning of SNR evolution, the
surface brightness map is very like the Sedov one (Fig.~\ref{surf-map}), but
with age the differences become more 
considerable when the majority  of emission arises from more dense regions.
The brightness contrast $S_{\rm max}/S_{\rm min}$ may increase with time
up to $\sim 10^{5}$ ($S_{\rm max}$ and $S_{\rm min}$ are the values 
of both the biggest and 
the smallest maxima in distribution of surface brightness). \par

   In order to interpret an observation, it is necessary to take into account
that projection effects also essentially affect the visible morphology of NSNR. 
Three cases of projection are shown. We can 
see that projection decreases real anisotropy and contrasts. For example, 
under condition of full disclosure of NSNR at the age $t=3300\ {\rm years}$ 
(Fig.~\ref{surf-map-bc}) 
$S_{\rm max}/S_{\rm min}=425$ but for inclination angle 
$45^{\rm o}$ this ratio is $S_{\rm max}/S_{\rm min}=75$. \par

It is interesting also to compare Fig.~\ref{surf-map} and
a bottom case of Fig.~\ref{surf-map-a} (when a visible shape of 
NSNR is
spherical as result of projection). One can see that even spherically 
symmetric Sedov-like observational distribution of surface brightness
does not guarantee an isotropic distribution of
parameters inside SNR i.e.,  uniformity of ISM density
and isotropy of explosion.

%%%%=== Fig ===%%%%%%%%%%%%%%%%%%%%%%%%%%%%%%%%%%%%%%%%%%%%%%%%%%%%%%%%%%%%%%%
\begin{figure}
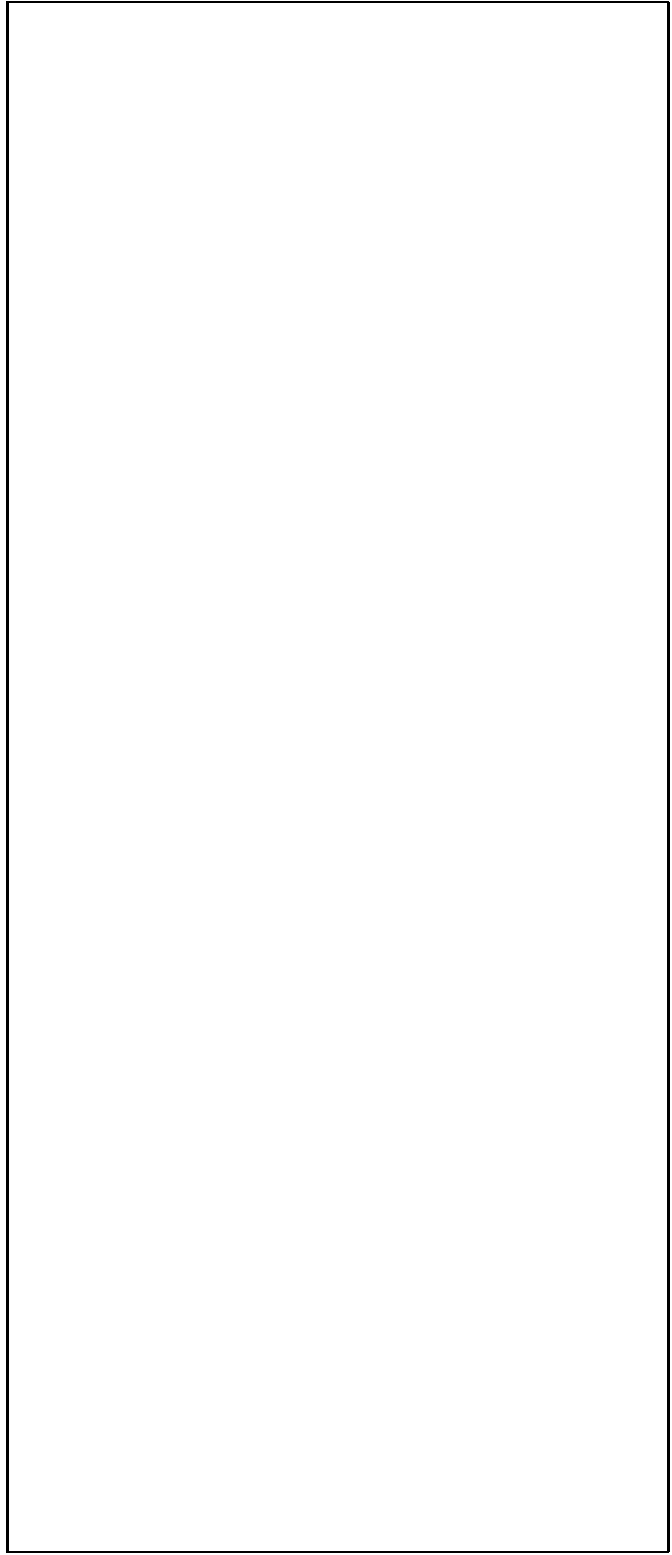

\picplace{20.60 cm}
%\epsfysize=20.6truecm
%\epsfbox{pc_12.eps}
\caption[]{
Distribution of surface brightness $S^{>0.1}$ 
of NSNR in medium with
power-law density distribution Eq.~(\ref{r^w-density}) in 
range $\varepsilon>0.1\ {\rm keV}$.
Model parameters are: $E_{\rm 51}=1,$ $n_{\rm H}^{\rm o}(0)=0.2\ {\rm cm^{-3}},$ 
$t=17800\ {\rm yrs},$ $r_{\rm o}=-17.5\ {\rm pc}$.
The NSNR inclination angle to the plane of the sky equals 
$0^{\rm o}$ (upper case), $45^{\rm o}$ (center case), 
and $90^{\rm o}$ (lower case). 
          }
\label{surf-r^w-a}
\end{figure}
%%%%%%%%%%%%%%%%%%%%%%%%%%%%%%%%%%%%%%%%%%%%%%%%%%%%%%%%%%%%%%%%%%%%%%%%%%%%%%
%%%%=== Fig ===%%%%%%%%%%%%%%%%%%%%%%%%%%%%%%%%%%%%%%%%%%%%%%%%%%%%%%%%%%%%%%%
\begin{figure*}
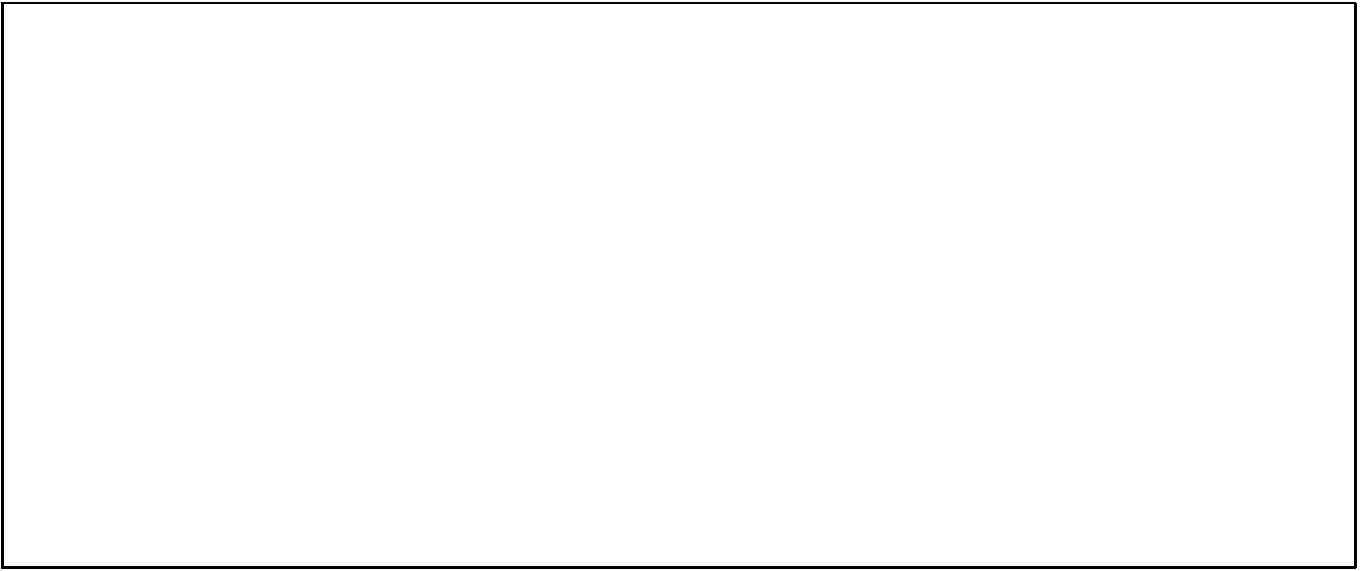

\picplace{7.5 cm}
%\epsfysize=7.5truecm
%\epsfbox{pc_13.eps}
\caption[]{
The same as in Fig.~\ref{surf-map-a} for range
$\varepsilon>4.5\ {\rm keV}$ and time $t=3300\ {\rm yrs}$ 
($\log(L_{\rm x}^{>4.5})=34.48$) and
$t=25500\ {\rm yrs}$ ($\log(L_{\rm x}^{>4.5})=32.58$).
          }
\label{surf-map-45keV}
\end{figure*}
%%%%%%%%%%%%%%%%%%%%%%%%%%%%%%%%%%%%%%%%%%%%%%%%%%%%%%%%%%%%%%%%%%%%%%%%%%%%%%

\subsubsection{NSNRs in a medium with power-law density distribution}

Fig.~\ref{surf-r^w-a} shows three projections of NSNR in power-law
density distribution (\ref{r^w-density}) with $E_{\rm 51}=1,$
$n_{\rm H}^{\rm o}(0)=0.2\ {\rm cm^{-3}},$ $t=1000\ {\rm yrs}$.
Corresponding NSNR characteristics are 
$\log(L_{\rm x}^{>0.1})=37.03,$ $M=268\ {\rm M_\odot},$ 
$T_{\rm ch}=3.5\cdot 10^6\ {\rm K}$. It may be compared with values 
calculated for NSNRs which evolve in media 
with different density distribution, 
to show that different ISM density distributions give similar integral
(surface-integrated) X-ray characteristics. For the same model parameters 
in case of SNR in uniform density, we obtain
$\log(L_{\rm x}^{>0.1})=36.75,$ $M=299\ {\rm M_\odot},$
$T_{\rm s}=3.1\cdot 10^6\ {\rm K}$ and for exponential NSNR:
$\log(L_{\rm x}^{>0.1})=37.08,$
$M=308\ {\rm M_\odot},$ $T_{\rm ch}=3.0\cdot 10^6\ {\rm K}.$ \par

As result of projection, the maximum of surface brightness 
does not lie close to the edges of NSNR as in
shell-like SNRs, but creates a compact region inside the visible projection.
Therefore, the projection effect in case of the NSNR elongated predominantly 
along the line of sight is one possible sources of apparent 
filled-centre SNRs, especially when the search for a pulsar has no result. \par

\subsection{Surface brightness in range $\varepsilon>4.5\ {\rm keV}$}

   Different photon energy ranges reveal different sensitivity to 
the non-uniformity
of ISM. Fig.~\ref{surf-map-45keV} shows the results of calculation of 
surface brightness of NSNR in an exponential density distribution 
(\ref{exp-density}) in range $\varepsilon>4.5\ {\rm keV}$. 
We may see that surface brightness contrast in this range is essentially 
smaller then in the wider range
$\varepsilon>0.1\ {\rm keV},$ where line emission dominates. 
So, for $t=3300\ {\rm years}$
(Fig.~\ref{surf-map-bc}) the maximal surface brightness
contrast is $S_{\rm max}/S_{\rm min}=425$ 
for $\varepsilon>0.1\ {\rm keV}$ and is
only $S_{\rm max}/S_{\rm min}=7$ for $\varepsilon>4.5\ {\rm keV}$. 
A contrast of surface 
brightness in range $>0.1\ {\rm keV}$ is mainly caused
by the contrast of the surrounding medium density distribution 
($S\propto\rho^{2}$),
but in range $>4.5\ {\rm keV},$ it weakly depends on density contrast 
($S\propto\rho^{1/2}$) (Hnatyk \& Petruk \cite{Hn-Pet-96}). \par

\subsection{Surface distribution of spectral index}

\begin{figure}
%\picplace{20.60 cm}
\epsfysize=20.6truecm
\epsfbox{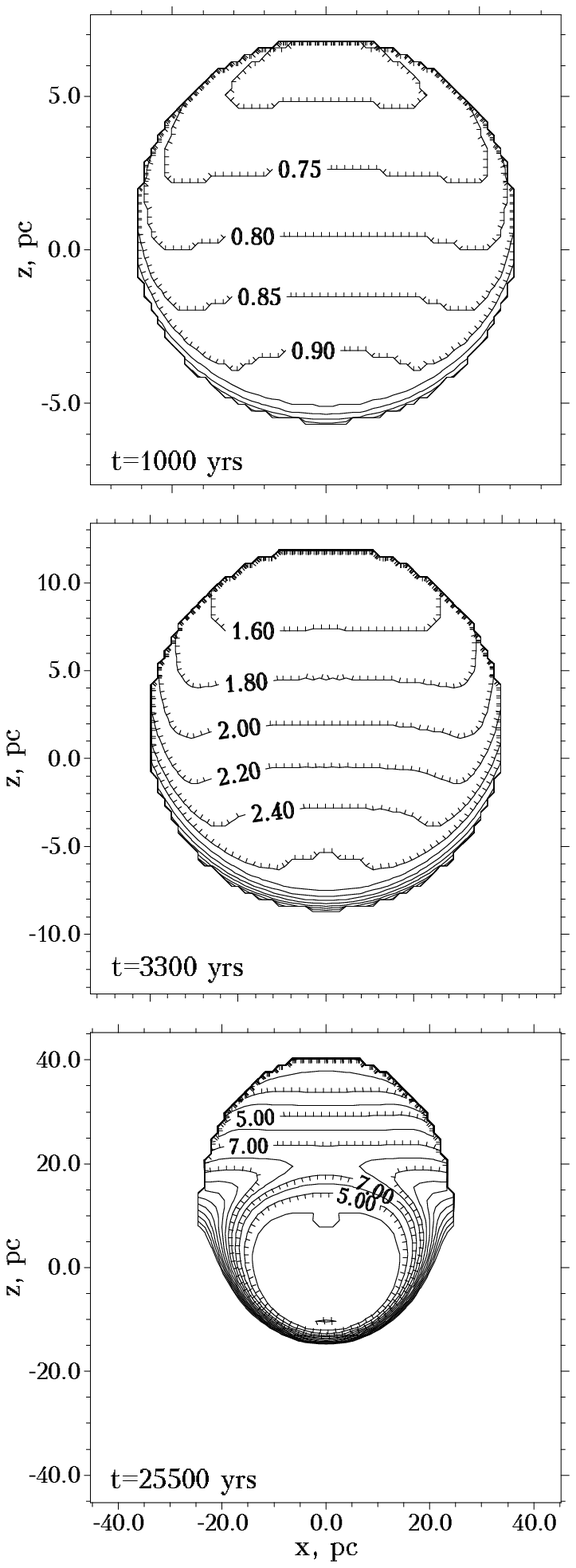}
\caption[]{
Surface distribution of spectral index $\alpha$ at 5 keV for NSNR in
exponential medium Eq.~(\ref{exp-density}) with $H=10\ {\rm pc},$ 
$E_{\rm 51}=1,$ $n_{\rm H}^{\rm o}(0)=0.2\ {\rm cm^{-3}}$ for time $t=1000\
{\rm yrs}$ (spectral index from total SNR $\alpha^{tot}(5\ {\rm keV})=0.86$), $t=3300\
{\rm yrs}$ ($\alpha^{tot}(5\ {\rm keV})=2.39$) and $t=25500\ {\rm yrs}$ ($\alpha^{tot}(5\
{\rm keV})=3.95$).
           }
\label{alpha-map-a}
\end{figure}

Plasma in different regions of NSNR is under different conditions
which influence emission. Therefore, spectra from different
NSNR regions must be different. The spectral index distribution for
NSNR in an exponential medium is shown in Fig.~\ref{alpha-map-a}. 
The contrast of values of spectral index increases with time but 
even for old adiabatic NSNR does not exceed a few times. Meantimes, 
the projection effects decrease both the index contrast and anisotropy 
of its distribution.

%%%%=== Fig ===%%%%%%%%%%%%%%%%%%%%%%%%%%%%%%%%%%%%%%%%%%%%%%%%%%%%%%%%%%%%%%%
%%%%%%%%%%%%%%%%%%%%%%%%%%%%%%%%%%%%%%%%%%%%%%%%%%%%%%%%%%%%%%%%%%%%%%%%%%%%%%

\subsection{Discussion}

We will summarize the results presented in this section. \par

1. The distribution of surface brightness of NSNR differs essentially from a 
spherical one. \par

2. The contrast of surface brightness increases with the age of the SNR
and may reach a few ten of thousands.  \par

3. Projection effects hide real contrasts of surface characteristics of
NSNR's emission. Observational morphology of NSNR depends essentially on
its orientation to the line of sight. \par

4. A harder X-ray range (e.g., $\varepsilon>4.5\ {\rm keV}$) is less 
sensitive to the influence of surrounding medium non-uniformity than a range 
which includes soft emission. This fact may be used for testing ISM 
non-uniformity. \par

5. The surface distribution of spectral index $\alpha$ may also be an
effective test for NSNR diagnostics.

%%%%%%%%%%%%%%%%%%%%%%%%%%%%%%%%%%%%%%%%%%%%%%%%%%%%%%%%%%%%%%%%%%%%%%%%%
%%%%======================Section V==================================%%%%
%%%%%%%%%%%%%%%%%%%%%%%%%%%%%%%%%%%%%%%%%%%%%%%%%%%%%%%%%%%%%%%%%%%%%%%%%
\section{Conclusions}

Large-scale non-uniformity of ISM density distribution in the SN 
neighborhood essentially affect the evolution of
SNR. The shape of SNR becomes essentially non-spherical and 
distributions of parameters inside the remnant, as viewed from the center 
of explosion, 
are strongly anisotropic. We have proposed a new approximate analytical
method for full hydrodynamical description of 3D point-like explosions
in non-uniform media with arbitrary density distribution i.e., for 
cases, when well-known self-similar Sedov solutions are inacceptable.
On the basis of it, we carry out the simulation of evolution of 2D 
non-spherical  
SNRs with special attention to their X-ray radiation. Since our aim in this 
paper was to 
investigate the role of density gradients in NSNR evolution, we restricted
ourself with the ionization equilibrium case in the 
calculation of X-ray radiation. The role of electron conductivity, 
nonequilibrium and nonequipartition effects will be investigated elsewhere.
\par
  
At first, we have investigated the shape of NSNR and brought out a remarkable 
fact of sphericity of {\it visible} shape of NSNR, even if the real deviation 
from sphericity is large. When $R_{\rm max}/R_{\rm min}<2,$ 
the observed shape will differ from spherical by less than $5\%$. 
Visual shape becomes noticeably non-spherical 
(more than 5\% of visual assymetry) only for surface density contrast of 
order 100. Projection effects on the sky plane decrease the real shape 
anisotropy.\par

    Therefore, we have calculated the equilibrium X-ray emission
characteristics of NSNR. Again, the total 
(surface-integrated) parameters of X-ray radiation (luminosity,
spectral index) are close to those in Sedov case with 
the same initial data. This is why many SNRs with apparently 
2D anisotropic
distribution of X-ray surface brightness are well described by the 
Sedov model.\par

   Contrary to previous cases of weak dependence of shape and
integral X-ray emission characteristics on density gradient, the 
distribution of X-ray emission characteristics along the surface of 
NSNR are very sensitive to an initial density distribution arround
SN. Moreover, contrast of X-ray surface brightness $S$ caused by density 
inhomogenity is most prominent, i.e., higher than 
density or 
temperature contrasts along the NSNR surface. For case of NSNR in 
Fig.~\ref{surf-map-bc} ($t=3300\ {\rm yrs}$), surface brightness contrast 
is $S_{\rm max}^{>0.1}/S_{\rm min}^{>0.1}=425,$ but contrast 
of shock density equals 10 and  
of shock temperature equals 4 only. \par

Therefore, the surface 
brightness maps give the most promising information concerning the
physical conditions inside and outside NSNR. It is important to note that 
typical contrasts of surface brightness caused by density inhomogenity 
(up to $\sim 10^5$) are considerably higher than those caused by
nonequilibrium effects. Therefore, the role of density gradient is 
dominant in interpretation of NSNR observations.\par
%%%%%%%%%%%%%%%%%%%%%%%%%%%%%%%%%%%%%%%%%%%%%%%%%%%%%%%%%%%%%%%%%%%%%%%%%
%%%%=======================Acknowledgements==========================%%%%
%%%%%%%%%%%%%%%%%%%%%%%%%%%%%%%%%%%%%%%%%%%%%%%%%%%%%%%%%%%%%%%%%%%%%%%%%
\begin{acknowledgements}
We are grateful to an anonymous referee for useful remarks and comments 
which helped us to improve the manuscript. 
\end{acknowledgements}
\medskip
%%%%%%%%%%%%%%%%%%%%%%%%%%%%%%%%%%%%%%%%%%%%%%%%%%%%%%%%%%%%%%%%%%%%%%%%%
%%%%=======================Appendix==================================%%%%
%%%%%%%%%%%%%%%%%%%%%%%%%%%%%%%%%%%%%%%%%%%%%%%%%%%%%%%%%%%%%%%%%%%%%%%%%
%%%%%%%%%%%%%%%%%%%%%%%%%%%%%%%%%%%%%%%%%%%%%%%%%%%%%%%%%%%%%%%%%%%%%%%%%
%%%%======================Appendix===================================%%%%
%%%%%%%%%%%%%%%%%%%%%%%%%%%%%%%%%%%%%%%%%%%%%%%%%%%%%%%%%%%%%%%%%%%%%%%%%

\appendix
\section{Appendix}
\subsection{Lagrangian form of hydrodynamical equations and shock
conditions}
\label{App-eqs}

   We use the set of hydrodynamical equations for the case of one-
dimensional adiabatic motion of nonviscous perfect gas in Lagrangian form
(Klimishin \cite{Klymyshyn})
\begin{equation}
\rho_t+{\rho^2 \over \rho^o}\Bigl({r\over a}\Bigr)^N u_a+ {{Nu\rho}
                      \over r}=0,
\end{equation}
\begin{equation}
u_t+{1\over \rho^o}\Bigl({r \over a}\Bigr)^N P_a=0,
\end{equation}
\begin{equation}
P_t-c^2\rho_t=0,
\end{equation}
\begin{equation}
r_t-u=0.
\end{equation}
Here gas pressure $P(a,t)$, its density $\rho(a,t)$, velocity $u(a,t)$
and Eulerian co-ordinate $r(a,t)$ are functions of Lagrangian co-ordinate, 
i.e., initial gas particle position $a$ and time $t$, $c=\sqrt{\gamma P
/\rho}$ is the adiabatic sound velocity, $\gamma$ is the adiabatic index.
Subscripts indicate partial derivatives with respect to corresponding 
variables, $\rho^o = \rho^o(a)$ is the initial density distribution.\par

 The continuity equation (A-4) may be written in the following form
\begin{equation}
\rho^o\cdot a^N \cdot da = \rho \cdot r^N \cdot dr
\end{equation}
or
\begin{equation}
r_a={\rho^o \over \rho} \Bigl( {a \over r} \Bigr)^N
\end{equation}
\par
    At front of a strong shock with trajectory $R=R(t),$ the following
conditions are satisfied
\begin{equation}
u^{\rm s}=\omega \dot R
\end{equation}
\begin{equation}
\rho^{\rm s}=(\rho^o)^{\rm s}/(1-\omega)
\end{equation}
\begin{equation}
P^{\rm s}=\omega(\rho^o)^{\rm s} \dot R^2
\end{equation}
\begin{equation}
r^{\rm s}=R
\end{equation}
where $\dot R=dR/dt$  is the shock velocity, superscript "s" corresponds to 
values of parameters at the shock front $a=R,\ \omega = 2/(\gamma +1)$. \par

\medskip
\subsection{Derivatives of functions of parameter distribution at the 
shock front}
\label{App-der}

    Equations (€-1) - (€-5) and shock conditions (€-6) - (€-9)
allow to find the values of arbitrary order partial derivatives of
hydrodynamical functions at the shock front using the law of shock motion
$R=R(t)$ (Gaffet \cite{Gaffet}).\par

   To find first derivatives of functions $\rho,\ P$ and
$u$ at front, we use equations (€-1) - (€-3), written for
$a=R$, and add three equations, resulting from differentiation of
boundary conditions
(€-6) - (€-8) along the shock trajectory by operator  $D/Dt =
(\partial/ \partial t) +\dot R (\partial/\partial a)$.  Solving
the obtained set of six equations for six unknown partial
derivatives at the shock front, we obtain expressions for 
$u^{\rm s}_a,\ u^{\rm s}_t,\
P^{\rm s}_a,\ P^{\rm s}_t,\ \rho^{\rm s}_a,\ \rho^{\rm s}_t$ 
(Hnatyk \& Petruk \cite{Hn-Pet-96}). \par

  From equation (€-5), we have now
\begin{equation}
\begin{array}{l} r^{\rm s}_a = 1-\omega, \\ \\ r^{\rm s}_t = u^{\rm s}, \\ \\ R r^{\rm s}_{aa}=
\omega(1-\omega) \bigl[ 3B+N(2-\omega)-m\bigr]
\end{array}
\end{equation}
where $B=R\ddot R/\dot R^2,\qquad m\equiv m(R)=-(dln\rho/dln a)^{\rm s}$.
\par

   To find second derivatives of hydrodynamical functions
at the shock front we differentiate equations (€-1) - (€-3) separately 
with respect to $a$ and  $t,$ at the shock front (a=R), what gives us six 
equations for nine unknown derivatives. Additional three equations are given 
by differentiation of shock conditions (€-6) - (€-8) by the operator
\begin{equation}
{D^2\over Dt^2}={\partial^2\over \partial t^2} +2\dot
R{\partial^2\over \partial a\partial t}+\dot R^2{\partial^2\over \partial
a^2}+\ddot R{\partial \over\partial a}.
\end{equation}
\par
   Solving the system of nine equations with nine unknowns, 
we obtain expressions for $u^{\rm s}_{at},\ u^{\rm s}_{aa},\ 
\rho^{\rm s}_{aa}$ (Hnatyk \& Petruk \cite{Hn-Pet-96}).  
From (€-5) we obtain
{\small 
\begin{equation} 
\begin{array}{l} 
R^2 r^{\rm s}_{aaa}=\omega
(1-\omega)\Bigl[ 3(7-5\omega)B^2+ \\ \qquad\quad +\bigl[
(-5\omega^2+4\omega +8)N+(4\omega -11)m\bigr]B+  \\ \qquad\quad
+\omega(2\omega ^2-7\omega+6)N^2+(\omega^2+\omega -4)Nm- \\ \qquad\quad
-\omega(2-\omega)N-(\omega-2)m^2+(2\omega-1)m+           \\ \qquad\quad
+(2\omega-1)m'+(6\omega-4)Q \Bigr] 
\end{array} 
\end{equation}} 
where $Q=R^2R^{(3)}/\dot R^3, \ \ \ m'=-dm/dlnR=-R\cdot dm/dR $.\par

   The time derivative $(r^{\rm s}_a)_t=0$, and for other ones we have
\begin{equation} {R\over \dot R}\cdot (R
r^{\rm s}_{aa})_t=\omega(1-\omega)\bigl[ 3(B+Q-2B^2)+m' \bigr] \end{equation}
{\small
\begin{equation}
\begin{array}{l}
{\displaystyle
{R\over \dot R}\cdot (R^2 r^{\rm s}_{aaa})_t=\omega(1-\omega)\Bigl[
3(7-5\omega)\cdot 2B(B+Q-2B^2)+                                       } \\
\quad +\bigl[ (-5\omega^2+4\omega+8)N+(4\omega-11)m\bigr]\cdot          \\
\qquad\qquad\qquad\qquad \cdot (B+Q-2B^2)-                                          \\
\quad -\bigl[(4\omega-11)B+(\omega^2+\omega-4)N+6(1-\omega)m\bigr]m'-   \\
\quad -(2\omega-1)\bigl[(2m+1)m'+m''\bigr]+                             \\
\quad +2(3\omega-2)\bigl[ 2Q+L-3BQ \bigr] \Bigr]
\end{array}
\!\!\!\!\!\!\!\!\!\!\!\!\!\!\!\!\!\!\!\!\!\!\!\!\!\!\!\!\!\!\!\!\!\!\!
\end{equation}
}
where
\begin{equation}
L={R^3R^{(4)}\over \dot R^4}={R\over \dot R}\dot Q-2Q+3BQ
\end{equation}
\par

   Derivatives from B and Q with respect to time are
\begin{equation}
\dot B={\dot R\over R}[B+Q-2B^2],\ \ \ \dot Q={\dot R\over R}[2Q+L-3BQ].
\end{equation}
In the self-similar case $[B+Q-2B^2]=0$,
${[2Q+L-3BQ]=0}$. \par
  According to the approximate formula for shock velocity (1) in the common
(non-similar) case
{\small
\begin{equation} B=k(m)[m-(N+1)] , \end{equation} \begin{equation}
Q=2k^2(m)[m-(N+1)]^2-k(m)[m'+m-(N+1)] , \end{equation} \begin{equation}
L=(4B-1)(B+Q-2B^2)+(3B-2)Q+k(m)m'' ,
\end{equation}
}
In the work of Hnatyk \& Petruk \cite{Hn-Pet-96}, the rest  derivatives
$u^{\rm s}_{tt},\ \rho^{\rm s}_{at},\ \rho^{\rm s}_{tt},\ P^{\rm s}_{aa},\ P^{\rm s}_{at},\
P^{\rm s}_{tt}$ are presented.\par

\medskip
\subsection{Approximation of the connection between
Lagrangian and Eulerian co-ordinates}
\label{App-r_o}

    The algorithm considered above allows to calculate partial 
derivatives of arbitrary order. We consider the case when an 
expansion of density, pressure and velocity into series are 
restricted by second order at shock front and first in the central 
region. Namely, if at time $t$ the shock position is $R(t)$, we
approximate a connection between Eulerian co-ordinate $r(a,t)$ and 
the Lagrangian one $a$ in the following way 
\begin{equation} {r(a,t)\over R(t)}={\Bigl({a\over R}\Bigr)}^x\cdot 
(1+\alpha\cdot \xi + \beta \cdot \xi^2 + \gamma \cdot \xi^3 + \delta \cdot 
\xi^4), 
\end{equation} 
where  $\xi =(R-a)/R$, $x=(\gamma - 1)/\gamma$, and for each sector 
parameters $\alpha, \beta, \gamma, \delta$ are choosen from the 
condition that partial derivatives at shock front 
$r^{s}_{a},\ r^{\rm s}_{aa} ,\ r^{\rm s}_{aaa}$ and 
in center of explosion $r^0_a$ correspond to their exact values:  
\begin{equation} 
\begin{array}{l} 
{\displaystyle 
\alpha = -r^{\rm s}_a +x, }\\ \\ 
{\displaystyle 
\beta ={1\over 2}\cdot \bigl(R 
r^{\rm s}_{aa}-2x\cdot r^{\rm s}_a + x(x+1)\bigr), }\\ \\ 
\gamma = {\displaystyle {1\over 6}\cdot } 
{\displaystyle 
\bigl(-R^2 r^{\rm s}_{aaa} +3x\cdot R r^{\rm s}_{aa}- }\\ \\
\quad\qquad -3x(1+x) \cdot r^{\rm s}_a+x(x+1)(x+2)\bigr), \\ \\ 
{\displaystyle 
 \delta = C - (1 + \alpha + \beta + \gamma).  }\\ \end{array} 
\!\!\!\!\!\!\!\!\!\!\!\!\!\!\!\!\!\!\!\!\!\!\!\!\!\!\!\!  
\end{equation} 
\par
In case of a uniform medium from the Sedov self-similar solution, it  
follows that in the central region at $r\approx 0$ the $r(a)$ dependence
is $r/R=C\cdot(a/R)^x$ and the factor $C$ is connected with central pressure
in the following way:  
\begin{equation} 
\begin{array}{l} 
{\displaystyle 
C = C_{\rm A} = \biggl({(3+N)^2\over 8}(\gamma 
  +1)\Bigl({\gamma+1 \over \gamma} \Bigr)^ {\gamma}\cdot }\\ \\ 
{\displaystyle 
\qquad\qquad\qquad \cdot R^{N+1} P(0,t){\alpha_{\rm A}(N,\gamma) 
\over E_o}\biggr)^{-1/(\gamma(N+1))}, } 
\end{array} 
\end{equation}
thus for $\gamma = 5/3$ in case $N=0$ (plane shock) $C_{\rm A}=1.1670$, and
in case  $N=2$ (spherical shock) $C_{\rm A}=1.0833$. \par

   In the more general case of anisotropic explosion with direction-dependent
energy release, $E=E(\Omega)$ and shock trajectory $R=R(\Omega,t),$ 
from (\ref{umova_C}) one obtains generalized condition for factor $C$: 
\begin{equation} \begin{array}{l} C(\Omega,t)= \\  
\\ {\displaystyle =C_{\rm A}\cdot\biggl({R(\Omega,t)^{N+1}\over V_{tot}(t)} 
\cdot {1\over N+1}\cdot {E_o\over E(\Omega)}\biggr)^{-1/(\gamma(N+1))}, 
 } 
\end{array} 
\end{equation} 
where $E_o=\int E(\Omega)\,d\Omega.$

%%%%%%%%%%%%%%%%%%%%%%%%%%%%%%%%%%%%%%%%%%%%%%%%%%%%%%%%%%%%%%%%%%%%%%%%%
%%%%======================References=================================%%%%
%%%%%%%%%%%%%%%%%%%%%%%%%%%%%%%%%%%%%%%%%%%%%%%%%%%%%%%%%%%%%%%%%%%%%%%%%

%%%%%%%%%%%%%%%%%%%%%%%%%%%%%%%%%%%%%%%%%%%%%%%%%%%%%%%%%%%%%%%%%%%%%%%%%%%%%%
%%%%======================================================================%%%%
%%%%%%%%%%%%%%%%%%%%%%%%%%%%%%%%%%%%%%%%%%%%%%%%%%%%%%%%%%%%%%%%%%%%%%%%%%%%%%
\end{document}